\documentclass[%
 reprint,
 amsmath,amssymb,
 aps,
]{revtex4-2}

\usepackage{multirow}
\usepackage{array}
\usepackage{amsmath}
\usepackage[colorlinks=true, allcolors=blue]{hyperref}

\usepackage{xcolor}
\usepackage{graphicx}
\usepackage{dcolumn}

\usepackage{bm}
\usepackage{comment}

\begin{document}


\title{Universal Model of Optical-Field Electron Tunneling from Two-Dimensional Materials}


\author{Yi Luo}
\email{yi$_$luo@sutd.edu.sg}

\author{H. Y. Yang}

\author{Yee Sin Ang}
\email{yeesin$_$ang@sutd.edu.sg}

\author{L. K. Ang}
\email{ricky$_$ang@sutd.edu.sg}

\affiliation{%
Science, Mathematics and Technology, Singapore University of Technology and Design, 8 Somapah Road, Singapore 487372, Singapore.
}%

\begin{abstract}
We develop analytical models of optical-field electron tunneling from the edge and surface of two-dimensional (2D) materials, including the effects of reduced dimensionality, non-parabolic energy dispersion, band anisotropy, quasi-time dependent tunneling and emission dynamics indueced by the laser field. We discover a universal scaling between the tunneling current density $J$ and the laser electric field $F$: In($J/|F|^{\beta})\propto1/|F|$ with $\beta = 3 / 2$ in the edge emission and $\beta = 1$ in the vertical surface emission, which both are distinctive from the traditional Fowler-Nordheim (FN) model of $\beta = 2$. The current density exhibits an unexpected high-field saturation effect due to the reduced dimensionality of 2D materials, which is completely different from the space-charge saturation commonly observed in traditional bulk materials. Our results reveal the dc bias as an efficient method in modulating the optical-field tunneling sub-optical-cycle emission characteristics. Importantly, our model is in excellent agreement with a recent experiment on graphene. Our findings offer a theoretical foundation for the understanding of optical-field tunneling emission from the 2D material system, which is useful for the development of 2D-material based optoelectronics and vacuum nanoelectronics
\end{abstract}

\maketitle
\textcolor{blue}{\textit{Introduction.---}}Laser-matter interaction offers the capability for the manipulation of electron excitation and dynamics at ultrashort timescale, such as high-order harmonic generation \cite{ghimire2011observation,liu2017high}, carrier interband transition \cite{schultze2014attosecond}, spontaneous radiation \cite{rakhmatov2020bright}, photoelectron emission \cite{hommelhoff2006ultrafast,ropers2007localized,wu2008nonequilibrium,kruger2011attosecond,pant2012ultrafast,dombi2013ultrafast,pant2013time,zhang2016ultrafast}, quantum coherent control of excitation states \cite{sirotti2014multiphoton,forster2016two,reutzel2019coherent,zhou2022unraveling} and many others.
Among them, 
laser-triggered photoemission from solids has gained considerable current attention, because of its crucial role in the development of high-resolution electron microscopy and diffraction \cite{sun2020direct,sciaini2011femtosecond}, free electron lasers \cite{grguravs2012ultrafast}, tabletop laser accelerators \cite{peralta2013demonstration}, coherent electron sources \cite{polyakov2013plasmon,jones2016laser} and quantum nano-vacuum electronics \cite{lin2017electric,zhang2017100,zhang2021space,zhou2021ultrafast}.

At high laser intensity, the photoemission is due to the optical-field electron tunneling process (or optical-field emission) \cite{bormann2010tip,kruger2012attosecond}, where the strong time-varying light field greatly suppresses the material interface potential barrier enabling electron tunneling in the vicinity of Fermi level (cf. Fig. 1).
In this non-perturbative field-driven near-transient tunneling regime, incident optical electric field can achieve the steering of electron motion at a sub-cycle temporal scale, such as the subfemtosecond tunneling \cite{rybka2016sub,ludwig2020sub} and attosecond electron pulse generation \cite{kim2023attosecond}, which makes optical-field tunneling emission is important for ultrafast optical-field-driven electronics \cite{arashida2022subcycle} and on-chip attosecond lightwave
science \cite{goulielmakis2007attosecond,bionta2021chip}.

Other than metallic nanostrucutres \cite{kruger2011attosecond,dombi2013ultrafast,bormann2010tip,herink2012field,park2012strong,putnam2017optical,xiong2020plasmon},
emerging two-dimensional (2D) materials \cite{higuchi2017light,son2018ultrafast,zhou2019ultrafast,heide2019interaction,sushko2021asymmetric}  are employed to study their optical-field photoemission as well, due to the distinctive electronic band structures and nanoscale sharpness which provide the huge localized field enhancement factor allowing sufficient electron emission in using low-power lasers. 
For instance, the most common 2D material graphene has the atomic-scale thickness resulting in significantly high field enhancement near the edge \cite{weiss2012graphene}, large electron mobility \cite{novoselov2004electric} and high thermal damage threshold \cite{david2014evaluating}, which explains graphene being used effectively to realize ultrafast optical-field electron tunneling emission \cite{son2018ultrafast}. 
Despite tremendous ongoing efforts to implement 2D materials as emitters for ultrafast applications, the theoretical description of 2D material electron emission relies heavily on the Fowler-Nordheim (FN) formalism (which was developed for the traditional bulk materials) that is fundamentally incompatible with the reduced dimensionality and electronic properties of 2D materials \cite{qin2011analytical,ang2017theoretical,ang2018universal,ang2019generalized,ang2021physics,chan2021thermal,chan2022field,ang2023universal}. The optical-field tunneling model of 2D materials thus remains an urgent and open question that has yet to be addressed.

In this work, we develop universal models for optical-field tunneling emission from the edge and surface of a wide class of 2D materials, which explicitly considers the reduction of dimensionality, non-parabolic energy dispersion of 2D material, quasi-time dependent tunneling and electron emission dynamics driven by ultrafast laser. 
We reveal a universal current-field scaling: In($J/|F|^{\beta})\propto1/|F|$ with $\beta = 3/2$ for the edge emission and $\beta$ = 1 for the vertical surface emission respectively, which are different from the conventional FN law of $\beta = 2$ \cite{fowler1928electron}.
A peculiar saturation of photocurrent at ultrahigh laser intensity is identified due to the reduced dimensionality. 
Importantly, our model is in an excellent agreement to a recent experiment \cite{son2018ultrafast} of optical-field emission from monolayer graphene over different combinations of dc and laser fields.
Our model provides a universal description of optical-field tunnelling phenomena for a wide variety of 2D materials and shall form an important foundation for the development of novel 2D-material based ultrafast optical-field-driven tunneling nano-vacuum optoelectronics and device engineering.

\textcolor{blue}{\textit{Theoretical formalism.---}}We consider a 2D material lying in the $x-y$ plane with a dc field $F_0$ under the illumination of a laser field $F_1(t)$ perpendicular to the edge. 
The optical-field-induced tunneling current density via the edge along $x$ direction is \cite{ang2019generalized,ang2018universal}
\begin{equation}
      J_{e}(t)=\frac{eg}{(2\pi)^2}\sum_{k_\perp^{(i)}}\int [v_x(\varepsilon_{\parallel})f_{FD}(\varepsilon_{\parallel})\Gamma(\varepsilon_x,t)]d^{2}\emph{\textbf{k}}_{\parallel},
\label{eq1}
\end{equation}
where $e$ is the electron charge, $g$ is the spin-valley degeneracy factor, $\emph{\textbf{k}}_{\parallel}=(k_x,k_y)$ and $\varepsilon_{\parallel}$ are the electron wavevector and energy component in the $x-y$ plane, respectively.
Here, $v_x(\varepsilon_{\parallel})=\hbar^{-1} \partial\varepsilon_{\parallel}/\partial k_x$ is the electron $x$-direction velocity ($\hbar$ is the reduced Planck constant), $f_{FD} (\varepsilon_{\parallel})$ is the zero-temperature Fermi-Dirac distribution function, $\Gamma(\varepsilon_x,t)$ is the time-dependent electron transmission probability through the edge at energy component $\varepsilon_x$ [cf. Eq. (\ref{eq3})], $k_\perp^{(i)}\;(i = 1, 2, 3, …)$ denotes the discrete bound states along the direction orthogonal to the $x-y$ plane, due to the confinement of electrons within the 2D plane, and $\sum_{k_\perp^{(i)}}$ denotes the summation of all these discrete bound states. 

Consider a 2D material with a general form of anisotroppic energy dispersion,
\begin{equation}
      \varepsilon_{\parallel}(k_{x},k_{y})=\left(\alpha_{l}k_{x}^{2}+\beta_{l}k_{y}^{2}\right)^{l/2},
\label{eq2}
\end{equation}
where $\alpha_{l}$ and $\beta_{l}$ are the material-dependent parameters along $x$ and $y$ directions respectively, and $l \in \mathbb{Z}^{>0}$. 
Using Eq. (\ref{eq2}), we derive $v_x(\varepsilon_{\parallel})=
\hbar^{-1}l\alpha_{l}k_{x}(\alpha_{l}k_{x}^{2}+\beta_{l}k_{y}^{2})^{(l-2)/2}$, and 
$f_{FD} (\varepsilon_{\parallel})=H(\varepsilon_{F}-\varepsilon_{\parallel})=H(\varepsilon_{F}^{2/l}-\alpha_{l}k_{x}^{2}-\beta_{l}k_{y}^{2})$, where $H(x)$ denotes the Heaviside function.

\begin{figure}[t]
\centering %
\includegraphics[width=0.35\textwidth]{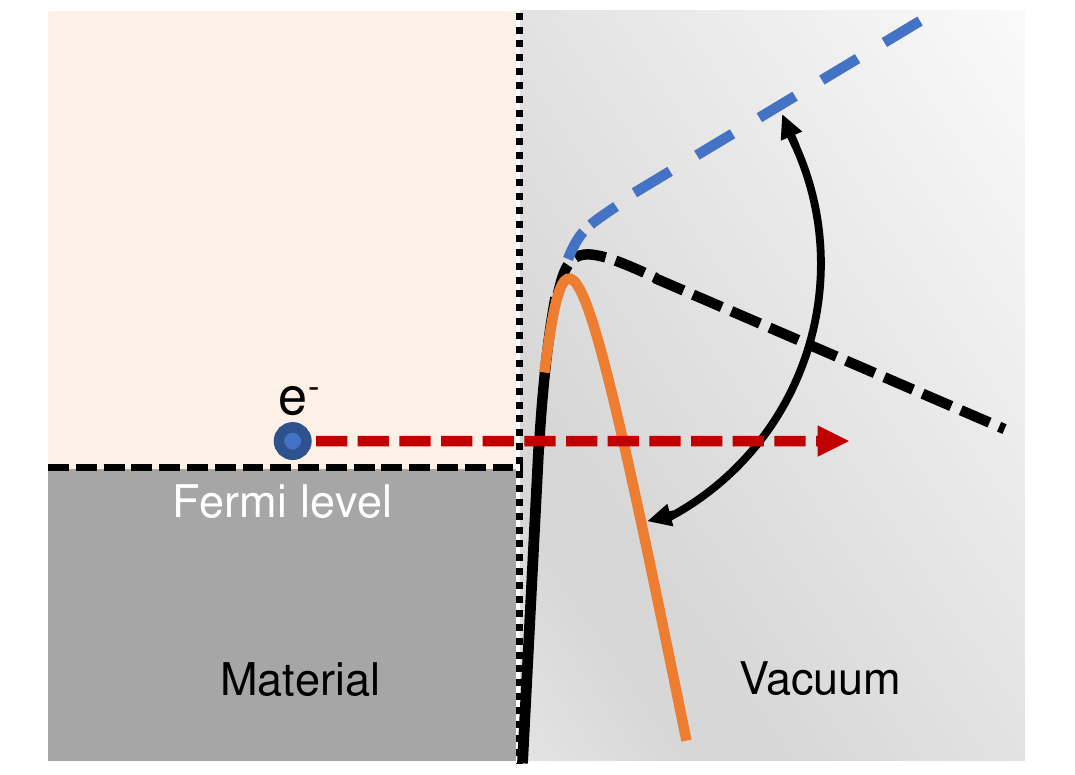} 
\caption{Schematic diagram for optical-field electron tunneling from the material surface with the actions of dc and optical fields. Under the illumination of high-intensity laser field, electron is liberated from the material-vacuum interface by tunneling through the time-varying vacuum potential barrier.}
\label{fig1}
\end{figure}

\begin{table*}[t]
\begin{center}
\caption{Energy dispersion (column 1), optical edge tunneling current density at practical field regime (column 2) and saturated edge current density at ultrahigh field regime (column 3) of general anisotropic 2D materials, monolayer graphene, highly doped multilayer black phosphorus, and $ABC$-stacked trilayer graphene. Here, $G=C_{2}C_{6}\frac{\Gamma(l/2)}{2\Gamma(3/2)\Gamma[(l+1)/2]}+C_{3}C_{7}\{1-\frac{\Gamma(l/2)}{2\Gamma(3/2)\Gamma[(l+1)/2]}\}$, $M=C_{1}\frac{\Gamma(l/2)}{2\Gamma(3/2)\Gamma[(l+1)/2]}+C_{4}\{1-\frac{\Gamma(l/2)}{2\Gamma(3/2)\Gamma[(l+1)/2]}\}$, $N=C_{5}C_{8}$ and $R=C_{5}$.}
\includegraphics[width=\textwidth]{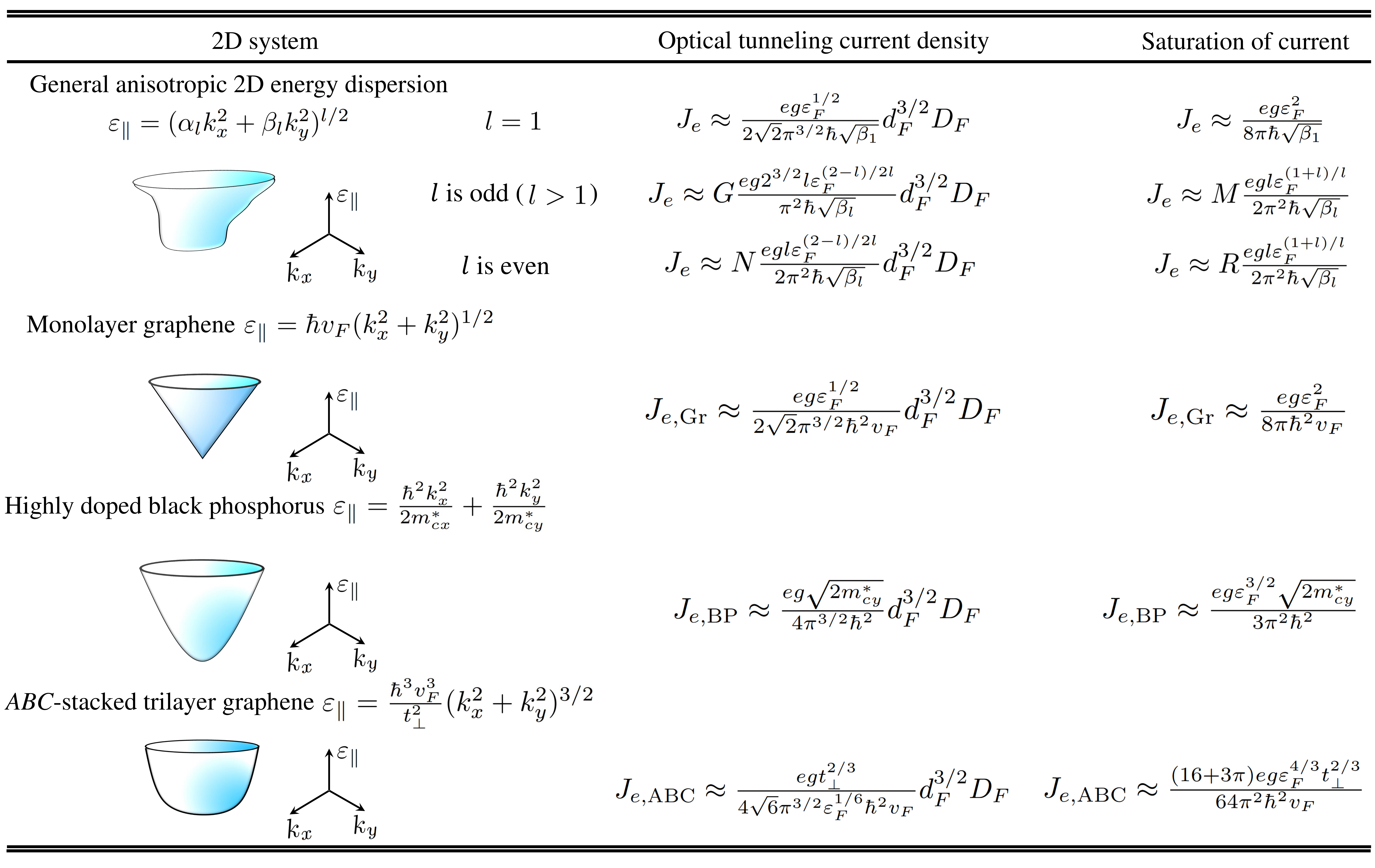}
\vspace*{-1cm}
\label{table1}
\end{center}
\end{table*}

In the optical high-field tunneling regime, incident optical field primarily behaves like an ac electric field that modulates the barrier, in which the effects of photoexcited nonthermal electrons can be ignored \cite{son2018ultrafast,wu2008nonequilibrium}.
Thus, the optical-field tunneling through the barrier resembles instantaneous field emission as a function of laser field, and the time-dependent transmission probability is approximated at quasi-static limit based on the Wentzel-Kramers-Brillouin (WKB) method, which gives
\begin{equation}
    \Gamma(\varepsilon_{x},t)=H\left[-F(t)\right] \times D_F(t) \times\exp\left(\frac{\varepsilon_{x}-\varepsilon_{F}}{d_F(t)}\right).
\label{eq3}
\end{equation}
Here, $F(t)$ is the summation of the dc field $F_0$ and the time-dependent optical field $F_1(t)$, and $H[-F(t)]$ denotes the Heaviside function to ensure electron emission occurs only during the negative field cycles. 
We omit $H[-F(t)]$ in the expressions below 
for simplicity.
The tunneling factor is $D_F(t)=\exp\left[-2W/3d_F(t)\right]$, where $d_F(t)=e\hbar|F(t)|/2\sqrt{2mW}$, $W$ is the work function, $\varepsilon_F$ is the Fermi energy, and $m$ is the free-electron mass. 
Substituting Eq. (\ref{eq3}) into Eq. (\ref{eq1}) yields the optical-field edge tunneling current density, 
$J_{e}=A_l \times \alpha_{l}^{l/2} \times K(l)$, where 
$A_l=egl\varepsilon_{F}^{1/l}D_{F}/2\pi^2\hbar\sqrt{\beta_{l}}$, and $K(l)$ is a complex integral as shown in Eq. (S4) in  the Supplemental Materials (SM).
In optical-field tunneling, the emission electrons are mostly located near the Fermi level: $\alpha_{l}^{l/2}k_{x}^{l} \to \varepsilon_{F}$, thus $\sqrt{\varepsilon_{F}^{2/l}/\alpha_{l}k_{x}^2-1} \ll 1$. Under this approximation, the tunneling current $J_{e}$ can be analytically solved as (see Sec. I in the SM for details)

\begin{table*}[t]
\begin{center}
\caption{Density of state (column 1), optical surface tunneling current density at practical intensity regime $d_F \ll \epsilon_{F}$ (column 2) and saturated surface current density at ultrahigh field regime $d_F\gg \epsilon_{F}$ (column 3) of general anisotropic 2D materials, monolayer graphene, highly doped multilayer black phosphorus, and $ABC$-stacked trilayer graphene.}

\includegraphics[width=\textwidth]{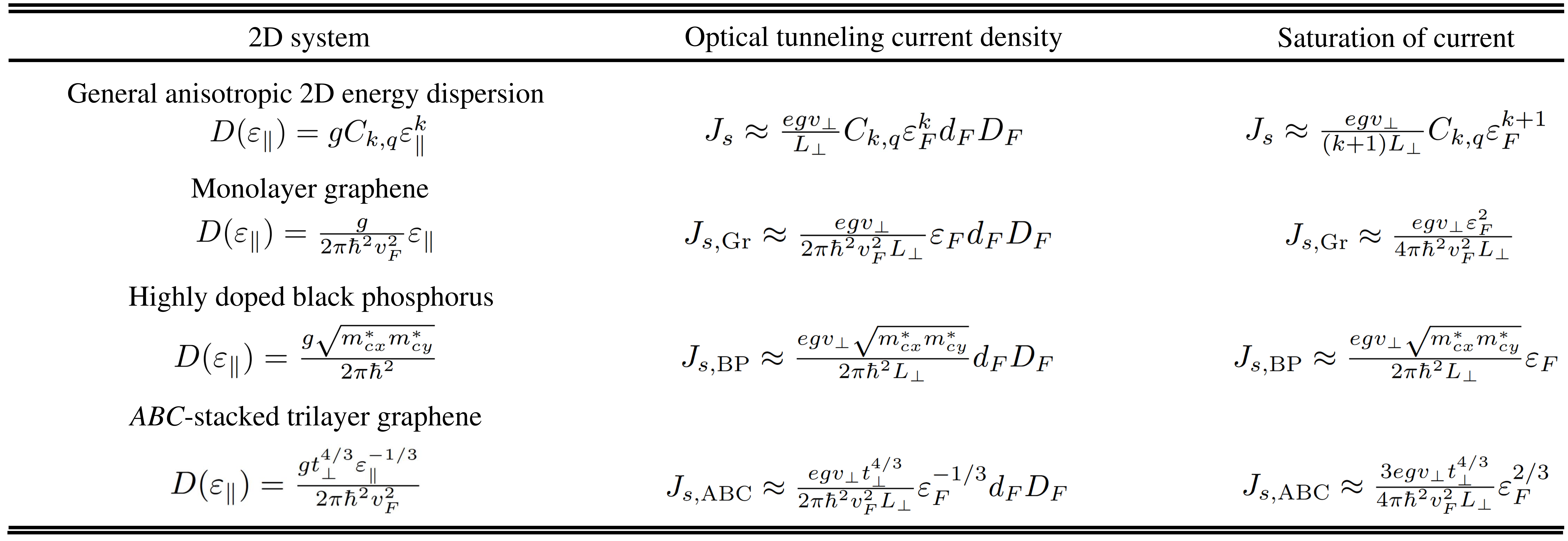}
\vspace*{-1cm}
\label{table2}
\end{center}
\end{table*}

\begin{widetext}
\begin{equation}
J_{e}\approx \begin{cases}
\vspace{5mm}
B_{1}d_{F}[I_{1}(\frac{\varepsilon_{F}}{d_{F}})+L_{1}(\frac{\varepsilon_{F}}{d_{F}})] \text{exp}(-\frac{\varepsilon_{F}}{d_{F}}), & \text{$l=1$}\\

B_{2}[C_{1}\hspace{0.08cm}_{l-1}F_{l}(\zeta_{1},\xi_{1},\frac{\varepsilon_{F}^{2}}{4d_{F}^2})+C_{2}\frac{\varepsilon_{F}}{d_{F}}\hspace{0.05cm}_{l}F_{l+1}(\zeta_{2},\xi_{2},\frac{\varepsilon_{F}^{2}}{4d_{F}^2})]\text{exp}(-\frac{\varepsilon_{F}}{d_{F}})+\\ \vspace{5mm}B_{3}[C_{3}\frac{\varepsilon_{F}}{d_{F}}\hspace{0.05cm}_{l-1}F_{l}(\zeta_{3},\xi_{3},\frac{\varepsilon_{F}^{2}}{4d_{F}^2})+C_{4}\hspace{0.08cm}_{l}F_{l+1}(\zeta_{4},\xi_{4},\frac{\varepsilon_{F}^{2}}{4d_{F}^2})]\text{exp}(-\frac{\varepsilon_{F}}{d_{F}}), & \text{$l$ is odd (= 3, 5, 7, ...)}\\

B_{4}C_{5}\hspace{0.08cm}_{l/2}F_{l/2}(\zeta_{5},\xi_{5},\frac{\varepsilon_{F}}{d_{F}}) \text{exp}(-\frac{\varepsilon_{F}}{d_{F}}), & \text{$l$ is even (= 2, 4, 6, ...)}

          \end{cases}
\label{eq4}
\end{equation}
\end{widetext}
where $B_{1}=A_1\pi/2$, $A_1$ is $A_l$ with $l=1$, $I_{1}$ is the modified Bessel function of the first kind, $L_{1}$ is the modified Struve function, $B_{2}=A_l\varepsilon_{F}\frac{\Gamma(l/2)}{2\Gamma(3/2)\Gamma[(l+1)/2]}$, 
$B_{3}=A_l\varepsilon_{F}\{1-\frac{\Gamma(l/2)}{2\Gamma(3/2)\Gamma[(l+1)/2]}\}$, $\Gamma(x)$ is the complete gamma function, 
$B_{4}=A_{l}\varepsilon_{F}$, 
$C_{1}, C_{2}, C_{3}, C_{4}$ and $C_{5}$ (including $C_{6}$, $C_{7}$ and $C_{8}$ below) all are $l$-dependent constants, and $_pF_q$ is the generalized hypergeometric function. 
The details of the constants and other parameters can be found in Sec. I in the SM.

The universal analytical $J_{e}$ [Eq. (\ref{eq4})] is characterized for three classes of the 2D materials [$l$ = 1, odd $l$ $(> 1)$ and even $l$], which shows good agreements with the numerical results from Eq. (S4) (cf. Fig. S1).
Equation (\ref{eq4}) also indicates that $J_{e}$ is independent of the material parameter $\alpha_{l}$, and thus a possible universal edge emission scaling.
At the practical field regime ($d_F \ll \varepsilon_{F}$) and the ultrahigh field regimes ($d_F\gg \varepsilon_{F}$), the analytical results for the universal model and three representative 2D systems are summarized in Table \ref{table1} (see Sec. II in the SM for detailed derivations). The validity of these analytical equations are verified in Fig. S2 in the SM.

Intriguingly, at practical intensity regime (column 2 of Table \ref{table1}), the prefactor ($d_{F}\propto|F|$) of the field strength exhibits an unconventional universal scaling of $J_{e} \propto |F|^{\beta}$ with $\beta = 3/2$ for $all$ $l$, which is in stark contrast to $\beta = 2$ of the conventinoal FN law \cite{fowler1928electron} derived for traditional 3D bulk materials.
Interestingly, the same universal $\beta = 3/2$ scaling has also been reported in the current-temperature scaling of the thermionic emission for lateral 2D-material-based Schottky contacts \cite{ang2018universal}. 
At ultrahigh field regime (column 3 of Table \ref{table1}), $J_{e}$ saturates and becomes independent of field strength $|F|$ (or $d_F$). 
Such saturation reveals a dramatic consequence of the reduced dimensional of 2D material, in which the emission current saturates due to limited availability of electrons. 
Such a source-limited emission behavior \cite{chua2021absence} is completely different from the space-charge limitation of traditional bulk materials \cite{zhang2017100,zhang2021space}, and shall offer a distinctive high-field transport signature of 2D-material-based electron optical-field emitters.

\setlength{\parskip}{0 pt}
\textcolor{blue}{\textit{Optical-field tunenling from 2D material surface.---}}We now consider the case of optical-field tunneling of electrons that occurs vertically from the plane of the 2D materials. 
In this case, the dc field $F_0$ and high-intensity optical field $F_1(t)$ are considered to be perpendicular to the 2D surface. 
During this out-of-plane tunneling process, carrier scattering effects, such as electron-electron \cite{meshkov1986tunneling} and defects scatterings \cite{chandni2016signatures}, violates the conservation of in-plane momentum 
$\emph{\textbf{k}}_{\parallel}$, which leads to the coupling between $\emph{\textbf{k}}_{\parallel}$ and $k_\perp^{(i)}$. Thus, the vertical optical-field tunneling current density from 2D material surface is expressed as 
\cite{ang2018universal},
\begin{equation}
    J_s(t)=\frac{eg}{(2\pi)^2}\sum_{k_\perp^{(i)}}\frac{v_\perp^{(i)}(k_\perp^{(i)})}{L_\perp} \int [f_{FD}(\varepsilon_{\parallel})\Gamma(\varepsilon_{\parallel},\varepsilon_\perp^{(i)},t)]d^{2}\emph{\textbf{k}}_{\parallel},
\label{eq8}
\end{equation}
where $L_\perp$ is the thickness of 2D material and $v_\perp^{(i)}(k_\perp^{(i)})$ 
is the cross-plane group velocity of the electron with discrete energy state $\varepsilon_\perp^{(i)}(k_\perp^{(i)})$. 
Consider the surface tunneling from one subband near the Fermi level, Eq. (\ref{eq8}) becomes
\begin{equation}
      J_s(t)=\frac{ev_\perp}{L_\perp}D_{F}(t) \int_{0}^{\varepsilon_{F}} \{D(\varepsilon_{\parallel})\text{exp}[(\varepsilon_{\parallel}-\varepsilon_{F})/d_{F}(t)]\} d{\varepsilon_{\parallel}},
\label{eq9}
\end{equation}
where $D(\varepsilon_{\parallel})$ is the density of state (DOS), and $D_F(t)$ and $d_F(t)$ are defined in Eq. (\ref{eq3}).

\renewcommand{\figurename}{FIG.}
\renewcommand{\thefigure}{\arabic{figure}}
\begin{figure*}
\centering
\includegraphics[width=\textwidth]{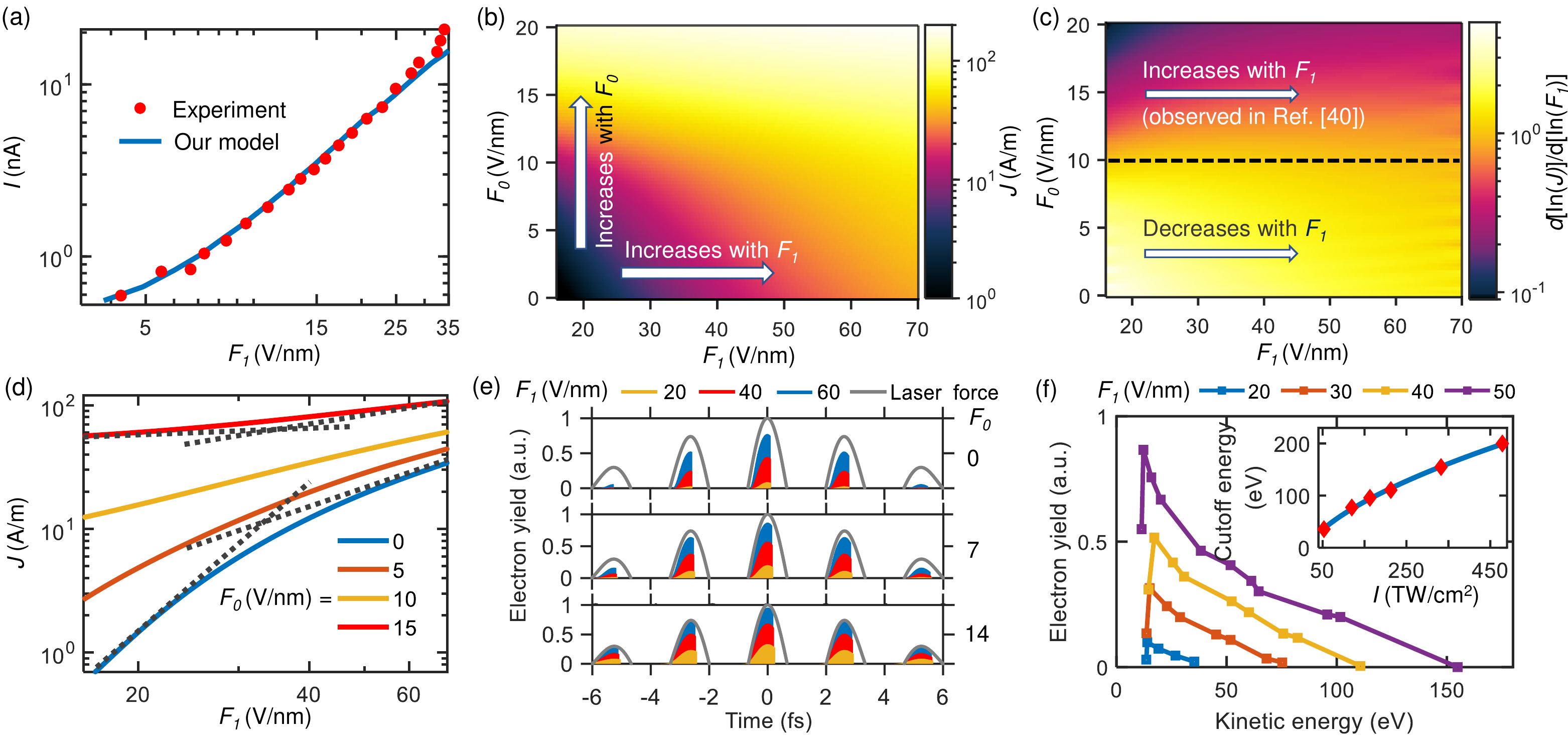}
\caption{
Calculated optical-field emission current with the exclusion of rescattered electrons from monolayer graphene’s edge. (a) Verification with the experimentally measured photocurrent $I$ as a function of laser field $F_1$ reported in \cite{son2018ultrafast}. 
Red circles are the experimental results. 
Solid blue line is the calculated result based on our developed model with dc and laser field
enhancement factors of about 146 and 22.
(b) Photocurrent density $J$ and (c) its power dependence on laser field $d[\text{In}(J)]/d[\text{In}(F_{1})]$ as a function of laser field $F_1$ and dc field $F_0$. 
(d) Photocurrent density $J$ as a function of laser field $F_1$ at dc field $F_0 = 0$, 5, 10, and 15 V/nm.
The dotted lines display the change of power dependence of $J$ on $F_1$ from the decrease to increase when $F_{0}$ varies from the small (= 0) to large (= 15 V/nm) value, which is reflected in (c).
(e) Time-dependent optical-field emission yield for different laser fields $F_1$ (filled yellow lines denote $F_1 = 20$ V/nm, red lines denote $F_1 = 40$ V/nm and blue lines denote $F_1 = 60$ V/nm) when the dc field $F_0 = 0$ (top), $F_0 = 7$ V/nm (middle) and $F_0 = 14$ V/nm (bottom). 
The grey curves denote the laser force on electrons. 
(f) Kinetic energy spectra of emitted electrons (far away from the edge of graphene) for different laser fields $F_1$. Inset in (f) shows the cutoff kinetic energy of photoelectron as a function of laser intensity $I$. In (f), the dc field is chosen to be zero.
Here, the work function and Fermi energy of graphene are set to be $W = 4.5$ eV and $\varepsilon_{F} = 0.1$ eV, respectively. 
}
\label{fig2}
\end{figure*}

Employing a general 2D anisotropic DOS $D(\varepsilon_{\parallel})=g \times C_{k,q} \times \varepsilon_{\parallel}^{k}$ with $C_{k,q}$ being material-dependent coefficients and $k,q \in \mathbb{Z}^{\geq0}$ \cite{ang2023universal}, Eq. (\ref{eq9}) yields a generalized analytical form 
\begin{equation}
      J_s(t)=\frac{egv_\perp}{L_\perp}D_{F}C_{k,q}(-d_{F})^{k+1}U\text{exp}(-\frac{\varepsilon_{F}}{d_{F}}),
\label{eq10}
\end{equation}
where $U=\Gamma(1+k)-\Gamma(1+k,-\varepsilon_{F}/{d_{F}})$, and $\Gamma(s)$ and $\Gamma(s,x)$ are the complete gamma function and the upper incomplete gamma function respectively. 
By taking the Taylor series expansion of $\Gamma(1+k,-\varepsilon_{F}/{d_{F}})$ for $d_F \ll \varepsilon_{F}$ and $d_F\gg \varepsilon_{F}$, $J_{s}$ is solved under these two limits for the universal formula and three different materials in Table \ref{table2} (see Sec. III of the SM for more details).

The vertical surface optical-field emission current all converge to an universal scaling of In($J_{s}/|F|^{\beta})\propto1/|F|$ with $\beta = 1$ at the practical field regime, which is different from $\beta=3/2$ of edge emission presented in Table \ref{table1} and $\beta=2$ of the classical FN law.
The current also saturates at the ultrahigh intensity due to the reduced dimensionality. 
The validity of these analytical solutions is shown in Fig. S3 in the SM.
By comparing Figs. S2 and S3, it is found under the same illumination condition, vertical surface optical-field tunneling generates significantly higher emission current than the edge tunneling.
Notably the analytical formulas in this work can be directly employed to study dc field emission by replacing $F(t)$ with dc field $F_{0}$.

\textcolor{blue}{\textit{Current-field characteristics.---}}We verify our model in comparison with a recently measured optical-field tunneling emission current from the edge of monolayer graphene \cite{son2018ultrafast} as shown in Fig. \ref{fig2}(a). 
Due to the significantly low backscattering efficiency on the carbon surface \cite{huang2006local}, the contributions from rescattered electrons are excluded in the calculations of Fig. {\ref{fig2}} (see Sec. IV of the SM for the method). 
Our model produces a good agreement with the experiment by using the dc and laser field enhancement factors of about 146 and 22 respectively, which are smaller than the reported values \cite{son2018ultrafast} based on the traditional bulk material FN law.

The optical-field edge emission current density $J$ from the graphene and its power dependence on the laser field are calculated as a function of laser field $F_1$ and dc field $F_0$ in Fig. \ref{fig2}(b) and \ref{fig2}(c), respectively. 
A Gaussian laser pulse is chosen, $F_1(t)=F_1 e^{-4\textnormal{In}2\,t^2/\tau^2} \cos(\omega t)$, where $F_1$ is the peak of field strength, $\tau$ is the pulse width, and $\omega$ is the angular frequency. 
The laser wavelength is 800 nm with a pulse width of $\tau$ = 8 fs (3 cycles) and the integration time [cf. Eq. (S40)] is from -30 to 30 fs. 
Increasing either laser intensity $F_1$ or dc bias $F_0$ will increase the optical-field emission [cf. Fig. \ref{fig2}(b)]. 
Interestingly, increasing $F_0$ changes the order of nonlinear power dependence of $J$ on $F_1$ from the decreasing to increasing trends [cf. Fig. \ref{fig2}(c) and dotted lines in Fig.\ref{fig2}(d)]. 
This transition occurs at around $F_{0} = 10$ V/nm and exists in the case without the consideration of emitted electron dynamics in the vacuum. 
Such behavior arises from the having higher electron tunneling probability through the barrier with stronger dc bias, thus allowing relatively more electrons to contribute in the high-intensity regime. 
When the relative number of emitted electrons is sufficiently large, the unconventional increasing exponential growth rate of current with laser intensity takes place [cf. the case with $F_0 > 10$ V/nm as shown in Fig. \ref{fig2}(c)]. 
This nonlinear order increment is also in agreement with experimental observations \cite{son2018ultrafast}.

\textcolor{blue}{\textit{Sub-cycle electron emission characteristics.---}}The time-dependent optical-field emission current $J_{e,\text{Gr}}(t)\delta(t)$ [cf. Eq. (S40)] in Fig. \ref{fig2}(e) shows that most electrons are emitted during the negative first half cycles due to the attainment of sufficient kinetic energy to escape the confinement of positive optical cycle. Beacsue of the reduced kinetic energy, parts of the electrons emitted in the negative second half cycles can be driven back to the edge by the deceleration of following positive cycle. 
This confirms the sub-optical-cycle emission dynamics in the optical-field emission regime. 
Increasing the dc field enlarges the rescattering time (cf. the sharp edge of emission yield), which implies that under the stronger dc acceleration, more electrons could escape the laser field and contribute to the net photocurrent.
Finally, we calculate the kinetic energy spectra at the position of around 25 nm away from the graphene edge for different laser field $F_1$ at zero dc field ($F_0$ = 0).
The extracted main spectrum peaks are plotted in Fig. \ref{fig2}(f). 
The cutoff kinetic energy representing the maximum kinetic energy of collected photoelectron as a function of laser intensity is depicted in the inset of Fig. \ref{fig2}(f).
The sublinear dependence of the cutoff energy on the laser intensity indicates a sub-cycle-timescale electron emission characteristics \cite{echternkamp2016strong} and the electron dynamic motion can be controlled by the optical field strength. 

\textcolor{blue}{\textit{Conclusion.---}}In summary, we have presented analytical and universal models for ultrafast optical-field tunneling emission of electrons from the edge and surface of general anisotropic 2D materials. 
Our model reports a universal scaling between the optical tunneling current density $J$ and laser electric field strength $F$ given by $\textnormal{In}(J/|F|^{\beta}) \propto 1/|F|$ with $\beta$ = 3/2 in the edge emission and $\beta$ = 1 in the vertical surface emission, which both are different from the traditional scaling of $\beta = 2$ based on the FN law. 
At ultrahigh laser field, we observe a saturation effect that the tunneling current no longer depends on the field, due to the reduction of dimensionality. 
These universal scaling and saturation are explicitly exhibited in three selected materials: monolayer grapehene, highly doped multilayer black phosphorus and $ABC$-stacked trilayer grapehene. 
Our calculation also well reproduces prior experimental measurement, and uncovers the dc induced modulation to the optical-field emission current and the power-dependence of photocurrent on laser field.
The developed analytical formulation provides a theoretical foundation for understanding a broad range of 2D materials in applications related to ultrafast optical-field-driven optoelectronics and field emission in vaccum nanoelectronics.
The model shall pave a way towards a generalized model combining thermal, field and photoemission \cite{jensen2007general} for 2D materials which is highly desirable for the design of 2D-material-based electron emitter.\\

This work is supported by the Singapore A*STAR IRG grant (A2083c0057). Y.S.A. is supported by the Singapore Ministry of Education (MOE) Academic Research Fund (AcRF) Tier 2 under Grant No. MOE-T2EP50221-0019.


\begin{thebibliography}{61}%
\makeatletter
\providecommand \@ifxundefined [1]{%
 \@ifx{#1\undefined}
}%
\providecommand \@ifnum [1]{%
 \ifnum #1\expandafter \@firstoftwo
 \else \expandafter \@secondoftwo
 \fi
}%
\providecommand \@ifx [1]{%
 \ifx #1\expandafter \@firstoftwo
 \else \expandafter \@secondoftwo
 \fi
}%
\providecommand \natexlab [1]{#1}%
\providecommand \enquote  [1]{``#1''}%
\providecommand \bibnamefont  [1]{#1}%
\providecommand \bibfnamefont [1]{#1}%
\providecommand \citenamefont [1]{#1}%
\providecommand \href@noop [0]{\@secondoftwo}%
\providecommand \href [0]{\begingroup \@sanitize@url \@href}%
\providecommand \@href[1]{\@@startlink{#1}\@@href}%
\providecommand \@@href[1]{\endgroup#1\@@endlink}%
\providecommand \@sanitize@url [0]{\catcode `\\12\catcode `\$12\catcode
  `\&12\catcode `\#12\catcode `\^12\catcode `\_12\catcode `\%12\relax}%
\providecommand \@@startlink[1]{}%
\providecommand \@@endlink[0]{}%
\providecommand \url  [0]{\begingroup\@sanitize@url \@url }%
\providecommand \@url [1]{\endgroup\@href {#1}{\urlprefix }}%
\providecommand \urlprefix  [0]{URL }%
\providecommand \Eprint [0]{\href }%
\providecommand \doibase [0]{https://doi.org/}%
\providecommand \selectlanguage [0]{\@gobble}%
\providecommand \bibinfo  [0]{\@secondoftwo}%
\providecommand \bibfield  [0]{\@secondoftwo}%
\providecommand \translation [1]{[#1]}%
\providecommand \BibitemOpen [0]{}%
\providecommand \bibitemStop [0]{}%
\providecommand \bibitemNoStop [0]{.\EOS\space}%
\providecommand \EOS [0]{\spacefactor3000\relax}%
\providecommand \BibitemShut  [1]{\csname bibitem#1\endcsname}%
\let\auto@bib@innerbib\@empty
\bibitem [{\citenamefont {Ghimire}\ \emph {et~al.}(2011)\citenamefont
  {Ghimire}, \citenamefont {DiChiara}, \citenamefont {Sistrunk}, \citenamefont
  {Agostini}, \citenamefont {DiMauro},\ and\ \citenamefont
  {Reis}}]{ghimire2011observation}%
  \BibitemOpen
  \bibfield  {author} {\bibinfo {author} {\bibfnamefont {S.}~\bibnamefont
  {Ghimire}}, \bibinfo {author} {\bibfnamefont {A.~D.}\ \bibnamefont
  {DiChiara}}, \bibinfo {author} {\bibfnamefont {E.}~\bibnamefont {Sistrunk}},
  \bibinfo {author} {\bibfnamefont {P.}~\bibnamefont {Agostini}}, \bibinfo
  {author} {\bibfnamefont {L.~F.}\ \bibnamefont {DiMauro}},\ and\ \bibinfo
  {author} {\bibfnamefont {D.~A.}\ \bibnamefont {Reis}},\ }\bibfield  {title}
  {\bibinfo {title} {Observation of high-order harmonic generation in a bulk
  crystal},\ }\href@noop {} {\bibfield  {journal} {\bibinfo  {journal} {Nature
  physics}\ }\textbf {\bibinfo {volume} {7}},\ \bibinfo {pages} {138} (\bibinfo
  {year} {2011})}\BibitemShut {NoStop}%
\bibitem [{\citenamefont {Liu}\ \emph {et~al.}(2017)\citenamefont {Liu},
  \citenamefont {Li}, \citenamefont {You}, \citenamefont {Ghimire},
  \citenamefont {Heinz},\ and\ \citenamefont {Reis}}]{liu2017high}%
  \BibitemOpen
  \bibfield  {author} {\bibinfo {author} {\bibfnamefont {H.}~\bibnamefont
  {Liu}}, \bibinfo {author} {\bibfnamefont {Y.}~\bibnamefont {Li}}, \bibinfo
  {author} {\bibfnamefont {Y.~S.}\ \bibnamefont {You}}, \bibinfo {author}
  {\bibfnamefont {S.}~\bibnamefont {Ghimire}}, \bibinfo {author} {\bibfnamefont
  {T.~F.}\ \bibnamefont {Heinz}},\ and\ \bibinfo {author} {\bibfnamefont
  {D.~A.}\ \bibnamefont {Reis}},\ }\bibfield  {title} {\bibinfo {title}
  {High-harmonic generation from an atomically thin semiconductor},\
  }\href@noop {} {\bibfield  {journal} {\bibinfo  {journal} {Nature Physics}\
  }\textbf {\bibinfo {volume} {13}},\ \bibinfo {pages} {262} (\bibinfo {year}
  {2017})}\BibitemShut {NoStop}%
\bibitem [{\citenamefont {Schultze}\ \emph {et~al.}(2014)\citenamefont
  {Schultze}, \citenamefont {Ramasesha}, \citenamefont {Pemmaraju},
  \citenamefont {Sato}, \citenamefont {Whitmore}, \citenamefont {Gandman},
  \citenamefont {Prell}, \citenamefont {Borja}, \citenamefont {Prendergast},
  \citenamefont {Yabana} \emph {et~al.}}]{schultze2014attosecond}%
  \BibitemOpen
  \bibfield  {author} {\bibinfo {author} {\bibfnamefont {M.}~\bibnamefont
  {Schultze}}, \bibinfo {author} {\bibfnamefont {K.}~\bibnamefont {Ramasesha}},
  \bibinfo {author} {\bibfnamefont {C.}~\bibnamefont {Pemmaraju}}, \bibinfo
  {author} {\bibfnamefont {S.}~\bibnamefont {Sato}}, \bibinfo {author}
  {\bibfnamefont {D.}~\bibnamefont {Whitmore}}, \bibinfo {author}
  {\bibfnamefont {A.}~\bibnamefont {Gandman}}, \bibinfo {author} {\bibfnamefont
  {J.~S.}\ \bibnamefont {Prell}}, \bibinfo {author} {\bibfnamefont
  {L.}~\bibnamefont {Borja}}, \bibinfo {author} {\bibfnamefont
  {D.}~\bibnamefont {Prendergast}}, \bibinfo {author} {\bibfnamefont
  {K.}~\bibnamefont {Yabana}}, \emph {et~al.},\ }\bibfield  {title} {\bibinfo
  {title} {Attosecond band-gap dynamics in silicon},\ }\href@noop {} {\bibfield
   {journal} {\bibinfo  {journal} {Science}\ }\textbf {\bibinfo {volume}
  {346}},\ \bibinfo {pages} {1348} (\bibinfo {year} {2014})}\BibitemShut
  {NoStop}%
\bibitem [{\citenamefont {Rakhmatov}\ \emph {et~al.}(2020)\citenamefont
  {Rakhmatov}, \citenamefont {Alizadehkhaledi}, \citenamefont {Hajisalem},\
  and\ \citenamefont {Gordon}}]{rakhmatov2020bright}%
  \BibitemOpen
  \bibfield  {author} {\bibinfo {author} {\bibfnamefont {E.}~\bibnamefont
  {Rakhmatov}}, \bibinfo {author} {\bibfnamefont {A.}~\bibnamefont
  {Alizadehkhaledi}}, \bibinfo {author} {\bibfnamefont {G.}~\bibnamefont
  {Hajisalem}},\ and\ \bibinfo {author} {\bibfnamefont {R.}~\bibnamefont
  {Gordon}},\ }\bibfield  {title} {\bibinfo {title} {Bright upconverted
  emission from light-induced inelastic tunneling},\ }\href@noop {} {\bibfield
  {journal} {\bibinfo  {journal} {Optics Express}\ }\textbf {\bibinfo {volume}
  {28}},\ \bibinfo {pages} {16497} (\bibinfo {year} {2020})}\BibitemShut
  {NoStop}%
\bibitem [{\citenamefont {Hommelhoff}\ \emph {et~al.}(2006)\citenamefont
  {Hommelhoff}, \citenamefont {Kealhofer},\ and\ \citenamefont
  {Kasevich}}]{hommelhoff2006ultrafast}%
  \BibitemOpen
  \bibfield  {author} {\bibinfo {author} {\bibfnamefont {P.}~\bibnamefont
  {Hommelhoff}}, \bibinfo {author} {\bibfnamefont {C.}~\bibnamefont
  {Kealhofer}},\ and\ \bibinfo {author} {\bibfnamefont {M.~A.}\ \bibnamefont
  {Kasevich}},\ }\bibfield  {title} {\bibinfo {title} {Ultrafast electron
  pulses from a tungsten tip triggered by low-power femtosecond laser pulses},\
  }\href@noop {} {\bibfield  {journal} {\bibinfo  {journal} {Physical review
  letters}\ }\textbf {\bibinfo {volume} {97}},\ \bibinfo {pages} {247402}
  (\bibinfo {year} {2006})}\BibitemShut {NoStop}%
\bibitem [{\citenamefont {Ropers}\ \emph {et~al.}(2007)\citenamefont {Ropers},
  \citenamefont {Solli}, \citenamefont {Schulz}, \citenamefont {Lienau},\ and\
  \citenamefont {Elsaesser}}]{ropers2007localized}%
  \BibitemOpen
  \bibfield  {author} {\bibinfo {author} {\bibfnamefont {C.}~\bibnamefont
  {Ropers}}, \bibinfo {author} {\bibfnamefont {D.}~\bibnamefont {Solli}},
  \bibinfo {author} {\bibfnamefont {C.}~\bibnamefont {Schulz}}, \bibinfo
  {author} {\bibfnamefont {C.}~\bibnamefont {Lienau}},\ and\ \bibinfo {author}
  {\bibfnamefont {T.}~\bibnamefont {Elsaesser}},\ }\bibfield  {title} {\bibinfo
  {title} {Localized multiphoton emission of femtosecond electron pulses from
  metal nanotips},\ }\href@noop {} {\bibfield  {journal} {\bibinfo  {journal}
  {Physical review letters}\ }\textbf {\bibinfo {volume} {98}},\ \bibinfo
  {pages} {043907} (\bibinfo {year} {2007})}\BibitemShut {NoStop}%
\bibitem [{\citenamefont {Wu}\ and\ \citenamefont
  {Ang}(2008)}]{wu2008nonequilibrium}%
  \BibitemOpen
  \bibfield  {author} {\bibinfo {author} {\bibfnamefont {L.}~\bibnamefont
  {Wu}}\ and\ \bibinfo {author} {\bibfnamefont {L.}~\bibnamefont {Ang}},\
  }\bibfield  {title} {\bibinfo {title} {Nonequilibrium model of ultrafast
  laser-induced electron photofield emission from a dc-biased metallic
  surface},\ }\href@noop {} {\bibfield  {journal} {\bibinfo  {journal}
  {Physical Review B}\ }\textbf {\bibinfo {volume} {78}},\ \bibinfo {pages}
  {224112} (\bibinfo {year} {2008})}\BibitemShut {NoStop}%
\bibitem [{\citenamefont {Kr{\"u}ger}\ \emph {et~al.}(2011)\citenamefont
  {Kr{\"u}ger}, \citenamefont {Schenk},\ and\ \citenamefont
  {Hommelhoff}}]{kruger2011attosecond}%
  \BibitemOpen
  \bibfield  {author} {\bibinfo {author} {\bibfnamefont {M.}~\bibnamefont
  {Kr{\"u}ger}}, \bibinfo {author} {\bibfnamefont {M.}~\bibnamefont {Schenk}},\
  and\ \bibinfo {author} {\bibfnamefont {P.}~\bibnamefont {Hommelhoff}},\
  }\bibfield  {title} {\bibinfo {title} {Attosecond control of electrons
  emitted from a nanoscale metal tip},\ }\href@noop {} {\bibfield  {journal}
  {\bibinfo  {journal} {Nature}\ }\textbf {\bibinfo {volume} {475}},\ \bibinfo
  {pages} {78} (\bibinfo {year} {2011})}\BibitemShut {NoStop}%
\bibitem [{\citenamefont {Pant}\ and\ \citenamefont
  {Ang}(2012)}]{pant2012ultrafast}%
  \BibitemOpen
  \bibfield  {author} {\bibinfo {author} {\bibfnamefont {M.}~\bibnamefont
  {Pant}}\ and\ \bibinfo {author} {\bibfnamefont {L.}~\bibnamefont {Ang}},\
  }\bibfield  {title} {\bibinfo {title} {Ultrafast laser-induced electron
  emission from multiphoton to optical tunneling},\ }\href@noop {} {\bibfield
  {journal} {\bibinfo  {journal} {Physical Review B}\ }\textbf {\bibinfo
  {volume} {86}},\ \bibinfo {pages} {045423} (\bibinfo {year}
  {2012})}\BibitemShut {NoStop}%
\bibitem [{\citenamefont {Dombi}\ \emph {et~al.}(2013)\citenamefont {Dombi},
  \citenamefont {Hörl}, \citenamefont {R{\'a}cz}, \citenamefont {M{\'a}rton},
  \citenamefont {Trügler}, \citenamefont {Krenn},\ and\ \citenamefont
  {Hohenester}}]{dombi2013ultrafast}%
  \BibitemOpen
  \bibfield  {author} {\bibinfo {author} {\bibfnamefont {P.}~\bibnamefont
  {Dombi}}, \bibinfo {author} {\bibfnamefont {A.}~\bibnamefont {Hörl}},
  \bibinfo {author} {\bibfnamefont {P.}~\bibnamefont {R{\'a}cz}}, \bibinfo
  {author} {\bibfnamefont {I.}~\bibnamefont {M{\'a}rton}}, \bibinfo {author}
  {\bibfnamefont {A.}~\bibnamefont {Trügler}}, \bibinfo {author}
  {\bibfnamefont {J.~R.}\ \bibnamefont {Krenn}},\ and\ \bibinfo {author}
  {\bibfnamefont {U.}~\bibnamefont {Hohenester}},\ }\bibfield  {title}
  {\bibinfo {title} {Ultrafast strong-field photoemission from plasmonic
  nanoparticles},\ }\href@noop {} {\bibfield  {journal} {\bibinfo  {journal}
  {Nano letters}\ }\textbf {\bibinfo {volume} {13}},\ \bibinfo {pages} {674}
  (\bibinfo {year} {2013})}\BibitemShut {NoStop}%
\bibitem [{\citenamefont {Pant}\ and\ \citenamefont
  {Ang}(2013)}]{pant2013time}%
  \BibitemOpen
  \bibfield  {author} {\bibinfo {author} {\bibfnamefont {M.}~\bibnamefont
  {Pant}}\ and\ \bibinfo {author} {\bibfnamefont {L.}~\bibnamefont {Ang}},\
  }\bibfield  {title} {\bibinfo {title} {Time-dependent quantum tunneling and
  nonequilibrium heating model for the generalized einstein photoelectric
  effect},\ }\href@noop {} {\bibfield  {journal} {\bibinfo  {journal} {Physical
  Review B}\ }\textbf {\bibinfo {volume} {88}},\ \bibinfo {pages} {195434}
  (\bibinfo {year} {2013})}\BibitemShut {NoStop}%
\bibitem [{\citenamefont {Zhang}\ and\ \citenamefont
  {Lau}(2016)}]{zhang2016ultrafast}%
  \BibitemOpen
  \bibfield  {author} {\bibinfo {author} {\bibfnamefont {P.}~\bibnamefont
  {Zhang}}\ and\ \bibinfo {author} {\bibfnamefont {Y.}~\bibnamefont {Lau}},\
  }\bibfield  {title} {\bibinfo {title} {Ultrafast strong-field photoelectron
  emission from biased metal surfaces: exact solution to time-dependent
  schr{\"o}dinger equation},\ }\href@noop {} {\bibfield  {journal} {\bibinfo
  {journal} {Scientific reports}\ }\textbf {\bibinfo {volume} {6}},\ \bibinfo
  {pages} {1} (\bibinfo {year} {2016})}\BibitemShut {NoStop}%
\bibitem [{\citenamefont {Sirotti}\ \emph {et~al.}(2014)\citenamefont
  {Sirotti}, \citenamefont {Beaulieu}, \citenamefont {Bendounan}, \citenamefont
  {Silly}, \citenamefont {Chauvet}, \citenamefont {Malinowski}, \citenamefont
  {Fratesi}, \citenamefont {V{\'e}niard},\ and\ \citenamefont
  {Onida}}]{sirotti2014multiphoton}%
  \BibitemOpen
  \bibfield  {author} {\bibinfo {author} {\bibfnamefont {F.}~\bibnamefont
  {Sirotti}}, \bibinfo {author} {\bibfnamefont {N.}~\bibnamefont {Beaulieu}},
  \bibinfo {author} {\bibfnamefont {A.}~\bibnamefont {Bendounan}}, \bibinfo
  {author} {\bibfnamefont {M.~G.}\ \bibnamefont {Silly}}, \bibinfo {author}
  {\bibfnamefont {C.}~\bibnamefont {Chauvet}}, \bibinfo {author} {\bibfnamefont
  {G.}~\bibnamefont {Malinowski}}, \bibinfo {author} {\bibfnamefont
  {G.}~\bibnamefont {Fratesi}}, \bibinfo {author} {\bibfnamefont
  {V.}~\bibnamefont {V{\'e}niard}},\ and\ \bibinfo {author} {\bibfnamefont
  {G.}~\bibnamefont {Onida}},\ }\bibfield  {title} {\bibinfo {title}
  {Multiphoton k-resolved photoemission from gold surface states with 800-nm
  femtosecond laser pulses},\ }\href@noop {} {\bibfield  {journal} {\bibinfo
  {journal} {Physical Review B}\ }\textbf {\bibinfo {volume} {90}},\ \bibinfo
  {pages} {035401} (\bibinfo {year} {2014})}\BibitemShut {NoStop}%
\bibitem [{\citenamefont {F{\"o}rster}\ \emph {et~al.}(2016)\citenamefont
  {F{\"o}rster}, \citenamefont {Paschen}, \citenamefont {Kr{\"u}ger},
  \citenamefont {Lemell}, \citenamefont {Wachter}, \citenamefont {Libisch},
  \citenamefont {Madlener}, \citenamefont {Burgd{\"o}rfer},\ and\ \citenamefont
  {Hommelhoff}}]{forster2016two}%
  \BibitemOpen
  \bibfield  {author} {\bibinfo {author} {\bibfnamefont {M.}~\bibnamefont
  {F{\"o}rster}}, \bibinfo {author} {\bibfnamefont {T.}~\bibnamefont
  {Paschen}}, \bibinfo {author} {\bibfnamefont {M.}~\bibnamefont {Kr{\"u}ger}},
  \bibinfo {author} {\bibfnamefont {C.}~\bibnamefont {Lemell}}, \bibinfo
  {author} {\bibfnamefont {G.}~\bibnamefont {Wachter}}, \bibinfo {author}
  {\bibfnamefont {F.}~\bibnamefont {Libisch}}, \bibinfo {author} {\bibfnamefont
  {T.}~\bibnamefont {Madlener}}, \bibinfo {author} {\bibfnamefont
  {J.}~\bibnamefont {Burgd{\"o}rfer}},\ and\ \bibinfo {author} {\bibfnamefont
  {P.}~\bibnamefont {Hommelhoff}},\ }\bibfield  {title} {\bibinfo {title}
  {Two-color coherent control of femtosecond above-threshold photoemission from
  a tungsten nanotip},\ }\href@noop {} {\bibfield  {journal} {\bibinfo
  {journal} {Physical Review Letters}\ }\textbf {\bibinfo {volume} {117}},\
  \bibinfo {pages} {217601} (\bibinfo {year} {2016})}\BibitemShut {NoStop}%
\bibitem [{\citenamefont {Reutzel}\ \emph {et~al.}(2019)\citenamefont
  {Reutzel}, \citenamefont {Li},\ and\ \citenamefont
  {Petek}}]{reutzel2019coherent}%
  \BibitemOpen
  \bibfield  {author} {\bibinfo {author} {\bibfnamefont {M.}~\bibnamefont
  {Reutzel}}, \bibinfo {author} {\bibfnamefont {A.}~\bibnamefont {Li}},\ and\
  \bibinfo {author} {\bibfnamefont {H.}~\bibnamefont {Petek}},\ }\bibfield
  {title} {\bibinfo {title} {Coherent two-dimensional multiphoton photoelectron
  spectroscopy of metal surfaces},\ }\href@noop {} {\bibfield  {journal}
  {\bibinfo  {journal} {Physical Review X}\ }\textbf {\bibinfo {volume} {9}},\
  \bibinfo {pages} {011044} (\bibinfo {year} {2019})}\BibitemShut {NoStop}%
\bibitem [{\citenamefont {Zhou}\ and\ \citenamefont
  {Zhang}(2022)}]{zhou2022unraveling}%
  \BibitemOpen
  \bibfield  {author} {\bibinfo {author} {\bibfnamefont {Y.}~\bibnamefont
  {Zhou}}\ and\ \bibinfo {author} {\bibfnamefont {P.}~\bibnamefont {Zhang}},\
  }\bibfield  {title} {\bibinfo {title} {Unraveling quantum pathways
  interference in two-color coherent control of photoemission with bias
  voltages},\ }\href@noop {} {\bibfield  {journal} {\bibinfo  {journal}
  {Physical Review B}\ }\textbf {\bibinfo {volume} {106}},\ \bibinfo {pages}
  {085402} (\bibinfo {year} {2022})}\BibitemShut {NoStop}%
\bibitem [{\citenamefont {Sun}\ \emph {et~al.}(2020)\citenamefont {Sun},
  \citenamefont {Sun}, \citenamefont {Bartles}, \citenamefont {Wozniak},
  \citenamefont {Williams}, \citenamefont {Zhang},\ and\ \citenamefont
  {Ruan}}]{sun2020direct}%
  \BibitemOpen
  \bibfield  {author} {\bibinfo {author} {\bibfnamefont {S.}~\bibnamefont
  {Sun}}, \bibinfo {author} {\bibfnamefont {X.}~\bibnamefont {Sun}}, \bibinfo
  {author} {\bibfnamefont {D.}~\bibnamefont {Bartles}}, \bibinfo {author}
  {\bibfnamefont {E.}~\bibnamefont {Wozniak}}, \bibinfo {author} {\bibfnamefont
  {J.}~\bibnamefont {Williams}}, \bibinfo {author} {\bibfnamefont
  {P.}~\bibnamefont {Zhang}},\ and\ \bibinfo {author} {\bibfnamefont {C.-Y.}\
  \bibnamefont {Ruan}},\ }\bibfield  {title} {\bibinfo {title} {Direct imaging
  of plasma waves using ultrafast electron microscopy},\ }\href@noop {}
  {\bibfield  {journal} {\bibinfo  {journal} {Structural Dynamics}\ }\textbf
  {\bibinfo {volume} {7}},\ \bibinfo {pages} {064301} (\bibinfo {year}
  {2020})}\BibitemShut {NoStop}%
\bibitem [{\citenamefont {Sciaini}\ and\ \citenamefont
  {Miller}(2011)}]{sciaini2011femtosecond}%
  \BibitemOpen
  \bibfield  {author} {\bibinfo {author} {\bibfnamefont {G.}~\bibnamefont
  {Sciaini}}\ and\ \bibinfo {author} {\bibfnamefont {R.~D.}\ \bibnamefont
  {Miller}},\ }\bibfield  {title} {\bibinfo {title} {Femtosecond electron
  diffraction: heralding the era of atomically resolved dynamics},\ }\href@noop
  {} {\bibfield  {journal} {\bibinfo  {journal} {Reports on Progress in
  Physics}\ }\textbf {\bibinfo {volume} {74}},\ \bibinfo {pages} {096101}
  (\bibinfo {year} {2011})}\BibitemShut {NoStop}%
\bibitem [{\citenamefont {Grgura{\v{s}}}\ \emph {et~al.}(2012)\citenamefont
  {Grgura{\v{s}}}, \citenamefont {Maier}, \citenamefont {Behrens},
  \citenamefont {Mazza}, \citenamefont {Kelly}, \citenamefont {Radcliffe},
  \citenamefont {D{\"u}sterer}, \citenamefont {Kazansky}, \citenamefont
  {Kabachnik}, \citenamefont {Tschentscher} \emph
  {et~al.}}]{grguravs2012ultrafast}%
  \BibitemOpen
  \bibfield  {author} {\bibinfo {author} {\bibfnamefont {I.}~\bibnamefont
  {Grgura{\v{s}}}}, \bibinfo {author} {\bibfnamefont {A.~R.}\ \bibnamefont
  {Maier}}, \bibinfo {author} {\bibfnamefont {C.}~\bibnamefont {Behrens}},
  \bibinfo {author} {\bibfnamefont {T.}~\bibnamefont {Mazza}}, \bibinfo
  {author} {\bibfnamefont {T.}~\bibnamefont {Kelly}}, \bibinfo {author}
  {\bibfnamefont {P.}~\bibnamefont {Radcliffe}}, \bibinfo {author}
  {\bibfnamefont {S.}~\bibnamefont {D{\"u}sterer}}, \bibinfo {author}
  {\bibfnamefont {A.}~\bibnamefont {Kazansky}}, \bibinfo {author}
  {\bibfnamefont {N.}~\bibnamefont {Kabachnik}}, \bibinfo {author}
  {\bibfnamefont {T.}~\bibnamefont {Tschentscher}}, \emph {et~al.},\ }\bibfield
   {title} {\bibinfo {title} {Ultrafast x-ray pulse characterization at
  free-electron lasers},\ }\href@noop {} {\bibfield  {journal} {\bibinfo
  {journal} {Nature Photonics}\ }\textbf {\bibinfo {volume} {6}},\ \bibinfo
  {pages} {852} (\bibinfo {year} {2012})}\BibitemShut {NoStop}%
\bibitem [{\citenamefont {Peralta}\ \emph {et~al.}(2013)\citenamefont
  {Peralta}, \citenamefont {Soong}, \citenamefont {England}, \citenamefont
  {Colby}, \citenamefont {Wu}, \citenamefont {Montazeri}, \citenamefont
  {McGuinness}, \citenamefont {McNeur}, \citenamefont {Leedle}, \citenamefont
  {Walz} \emph {et~al.}}]{peralta2013demonstration}%
  \BibitemOpen
  \bibfield  {author} {\bibinfo {author} {\bibfnamefont {E.}~\bibnamefont
  {Peralta}}, \bibinfo {author} {\bibfnamefont {K.}~\bibnamefont {Soong}},
  \bibinfo {author} {\bibfnamefont {R.}~\bibnamefont {England}}, \bibinfo
  {author} {\bibfnamefont {E.}~\bibnamefont {Colby}}, \bibinfo {author}
  {\bibfnamefont {Z.}~\bibnamefont {Wu}}, \bibinfo {author} {\bibfnamefont
  {B.}~\bibnamefont {Montazeri}}, \bibinfo {author} {\bibfnamefont
  {C.}~\bibnamefont {McGuinness}}, \bibinfo {author} {\bibfnamefont
  {J.}~\bibnamefont {McNeur}}, \bibinfo {author} {\bibfnamefont
  {K.}~\bibnamefont {Leedle}}, \bibinfo {author} {\bibfnamefont
  {D.}~\bibnamefont {Walz}}, \emph {et~al.},\ }\bibfield  {title} {\bibinfo
  {title} {Demonstration of electron acceleration in a laser-driven dielectric
  microstructure},\ }\href@noop {} {\bibfield  {journal} {\bibinfo  {journal}
  {Nature}\ }\textbf {\bibinfo {volume} {503}},\ \bibinfo {pages} {91}
  (\bibinfo {year} {2013})}\BibitemShut {NoStop}%
\bibitem [{\citenamefont {Polyakov}\ \emph {et~al.}(2013)\citenamefont
  {Polyakov}, \citenamefont {Senft}, \citenamefont {Thompson}, \citenamefont
  {Feng}, \citenamefont {Cabrini}, \citenamefont {Schuck}, \citenamefont
  {Padmore}, \citenamefont {Peppernick},\ and\ \citenamefont
  {Hess}}]{polyakov2013plasmon}%
  \BibitemOpen
  \bibfield  {author} {\bibinfo {author} {\bibfnamefont {A.}~\bibnamefont
  {Polyakov}}, \bibinfo {author} {\bibfnamefont {C.}~\bibnamefont {Senft}},
  \bibinfo {author} {\bibfnamefont {K.}~\bibnamefont {Thompson}}, \bibinfo
  {author} {\bibfnamefont {J.}~\bibnamefont {Feng}}, \bibinfo {author}
  {\bibfnamefont {S.}~\bibnamefont {Cabrini}}, \bibinfo {author} {\bibfnamefont
  {P.}~\bibnamefont {Schuck}}, \bibinfo {author} {\bibfnamefont
  {H.}~\bibnamefont {Padmore}}, \bibinfo {author} {\bibfnamefont {S.~J.}\
  \bibnamefont {Peppernick}},\ and\ \bibinfo {author} {\bibfnamefont {W.~P.}\
  \bibnamefont {Hess}},\ }\bibfield  {title} {\bibinfo {title}
  {Plasmon-enhanced photocathode for high brightness and high repetition rate
  x-ray sources},\ }\href@noop {} {\bibfield  {journal} {\bibinfo  {journal}
  {Physical review letters}\ }\textbf {\bibinfo {volume} {110}},\ \bibinfo
  {pages} {076802} (\bibinfo {year} {2013})}\BibitemShut {NoStop}%
\bibitem [{\citenamefont {Jones}\ \emph {et~al.}(2016)\citenamefont {Jones},
  \citenamefont {Becker}, \citenamefont {Luiten},\ and\ \citenamefont
  {Batelaan}}]{jones2016laser}%
  \BibitemOpen
  \bibfield  {author} {\bibinfo {author} {\bibfnamefont {E.}~\bibnamefont
  {Jones}}, \bibinfo {author} {\bibfnamefont {M.}~\bibnamefont {Becker}},
  \bibinfo {author} {\bibfnamefont {J.}~\bibnamefont {Luiten}},\ and\ \bibinfo
  {author} {\bibfnamefont {H.}~\bibnamefont {Batelaan}},\ }\bibfield  {title}
  {\bibinfo {title} {Laser control of electron matter waves},\ }\href@noop {}
  {\bibfield  {journal} {\bibinfo  {journal} {Laser \& Photonics Reviews}\
  }\textbf {\bibinfo {volume} {10}},\ \bibinfo {pages} {214} (\bibinfo {year}
  {2016})}\BibitemShut {NoStop}%
\bibitem [{\citenamefont {Lin}\ \emph {et~al.}(2017)\citenamefont {Lin},
  \citenamefont {Wong}, \citenamefont {Yang}, \citenamefont {Lau},
  \citenamefont {Tang},\ and\ \citenamefont {Zhang}}]{lin2017electric}%
  \BibitemOpen
  \bibfield  {author} {\bibinfo {author} {\bibfnamefont {J.}~\bibnamefont
  {Lin}}, \bibinfo {author} {\bibfnamefont {P.~Y.}\ \bibnamefont {Wong}},
  \bibinfo {author} {\bibfnamefont {P.}~\bibnamefont {Yang}}, \bibinfo {author}
  {\bibfnamefont {Y.}~\bibnamefont {Lau}}, \bibinfo {author} {\bibfnamefont
  {W.}~\bibnamefont {Tang}},\ and\ \bibinfo {author} {\bibfnamefont
  {P.}~\bibnamefont {Zhang}},\ }\bibfield  {title} {\bibinfo {title} {Electric
  field distribution and current emission in a miniaturized geometrical
  diode},\ }\href@noop {} {\bibfield  {journal} {\bibinfo  {journal} {Journal
  of Applied Physics}\ }\textbf {\bibinfo {volume} {121}},\ \bibinfo {pages}
  {244301} (\bibinfo {year} {2017})}\BibitemShut {NoStop}%
\bibitem [{\citenamefont {Zhang}\ \emph {et~al.}(2017)\citenamefont {Zhang},
  \citenamefont {Valfells}, \citenamefont {Ang}, \citenamefont {Luginsland},\
  and\ \citenamefont {Lau}}]{zhang2017100}%
  \BibitemOpen
  \bibfield  {author} {\bibinfo {author} {\bibfnamefont {P.}~\bibnamefont
  {Zhang}}, \bibinfo {author} {\bibfnamefont {{\'A}.}~\bibnamefont {Valfells}},
  \bibinfo {author} {\bibfnamefont {L.}~\bibnamefont {Ang}}, \bibinfo {author}
  {\bibfnamefont {J.}~\bibnamefont {Luginsland}},\ and\ \bibinfo {author}
  {\bibfnamefont {Y.}~\bibnamefont {Lau}},\ }\bibfield  {title} {\bibinfo
  {title} {100 years of the physics of diodes},\ }\href@noop {} {\bibfield
  {journal} {\bibinfo  {journal} {Applied Physics Reviews}\ }\textbf {\bibinfo
  {volume} {4}},\ \bibinfo {pages} {011304} (\bibinfo {year}
  {2017})}\BibitemShut {NoStop}%
\bibitem [{\citenamefont {Zhang}\ \emph {et~al.}(2021)\citenamefont {Zhang},
  \citenamefont {Ang}, \citenamefont {Garner}, \citenamefont {Valfells},
  \citenamefont {Luginsland},\ and\ \citenamefont {Ang}}]{zhang2021space}%
  \BibitemOpen
  \bibfield  {author} {\bibinfo {author} {\bibfnamefont {P.}~\bibnamefont
  {Zhang}}, \bibinfo {author} {\bibfnamefont {Y.~S.}\ \bibnamefont {Ang}},
  \bibinfo {author} {\bibfnamefont {A.~L.}\ \bibnamefont {Garner}}, \bibinfo
  {author} {\bibfnamefont {{\'A}.}~\bibnamefont {Valfells}}, \bibinfo {author}
  {\bibfnamefont {J.}~\bibnamefont {Luginsland}},\ and\ \bibinfo {author}
  {\bibfnamefont {L.}~\bibnamefont {Ang}},\ }\bibfield  {title} {\bibinfo
  {title} {Space--charge limited current in nanodiodes: Ballistic, collisional,
  and dynamical effects},\ }\href@noop {} {\bibfield  {journal} {\bibinfo
  {journal} {Journal of Applied Physics}\ }\textbf {\bibinfo {volume} {129}},\
  \bibinfo {pages} {100902} (\bibinfo {year} {2021})}\BibitemShut {NoStop}%
\bibitem [{\citenamefont {Zhou}\ \emph {et~al.}(2021)\citenamefont {Zhou},
  \citenamefont {Chen}, \citenamefont {Cole}, \citenamefont {Li}, \citenamefont
  {Li}, \citenamefont {Chen}, \citenamefont {Lienau}, \citenamefont {Li},\ and\
  \citenamefont {Dai}}]{zhou2021ultrafast}%
  \BibitemOpen
  \bibfield  {author} {\bibinfo {author} {\bibfnamefont {S.}~\bibnamefont
  {Zhou}}, \bibinfo {author} {\bibfnamefont {K.}~\bibnamefont {Chen}}, \bibinfo
  {author} {\bibfnamefont {M.~T.}\ \bibnamefont {Cole}}, \bibinfo {author}
  {\bibfnamefont {Z.}~\bibnamefont {Li}}, \bibinfo {author} {\bibfnamefont
  {M.}~\bibnamefont {Li}}, \bibinfo {author} {\bibfnamefont {J.}~\bibnamefont
  {Chen}}, \bibinfo {author} {\bibfnamefont {C.}~\bibnamefont {Lienau}},
  \bibinfo {author} {\bibfnamefont {C.}~\bibnamefont {Li}},\ and\ \bibinfo
  {author} {\bibfnamefont {Q.}~\bibnamefont {Dai}},\ }\bibfield  {title}
  {\bibinfo {title} {Ultrafast electron tunneling devices—from electric-field
  driven to optical-field driven},\ }\href@noop {} {\bibfield  {journal}
  {\bibinfo  {journal} {Advanced Materials}\ }\textbf {\bibinfo {volume}
  {33}},\ \bibinfo {pages} {2101449} (\bibinfo {year} {2021})}\BibitemShut
  {NoStop}%
\bibitem [{\citenamefont {Bormann}\ \emph {et~al.}(2010)\citenamefont
  {Bormann}, \citenamefont {Gulde}, \citenamefont {Weismann}, \citenamefont
  {Yalunin},\ and\ \citenamefont {Ropers}}]{bormann2010tip}%
  \BibitemOpen
  \bibfield  {author} {\bibinfo {author} {\bibfnamefont {R.}~\bibnamefont
  {Bormann}}, \bibinfo {author} {\bibfnamefont {M.}~\bibnamefont {Gulde}},
  \bibinfo {author} {\bibfnamefont {A.}~\bibnamefont {Weismann}}, \bibinfo
  {author} {\bibfnamefont {S.}~\bibnamefont {Yalunin}},\ and\ \bibinfo {author}
  {\bibfnamefont {C.}~\bibnamefont {Ropers}},\ }\bibfield  {title} {\bibinfo
  {title} {Tip-enhanced strong-field photoemission},\ }\href@noop {} {\bibfield
   {journal} {\bibinfo  {journal} {Physical review letters}\ }\textbf {\bibinfo
  {volume} {105}},\ \bibinfo {pages} {147601} (\bibinfo {year}
  {2010})}\BibitemShut {NoStop}%
\bibitem [{\citenamefont {Kr{\"u}ger}\ \emph {et~al.}(2012)\citenamefont
  {Kr{\"u}ger}, \citenamefont {Schenk}, \citenamefont {F{\"o}rster},\ and\
  \citenamefont {Hommelhoff}}]{kruger2012attosecond}%
  \BibitemOpen
  \bibfield  {author} {\bibinfo {author} {\bibfnamefont {M.}~\bibnamefont
  {Kr{\"u}ger}}, \bibinfo {author} {\bibfnamefont {M.}~\bibnamefont {Schenk}},
  \bibinfo {author} {\bibfnamefont {M.}~\bibnamefont {F{\"o}rster}},\ and\
  \bibinfo {author} {\bibfnamefont {P.}~\bibnamefont {Hommelhoff}},\ }\bibfield
   {title} {\bibinfo {title} {Attosecond physics in photoemission from a metal
  nanotip},\ }\href@noop {} {\bibfield  {journal} {\bibinfo  {journal} {Journal
  of Physics B: Atomic, Molecular and Optical Physics}\ }\textbf {\bibinfo
  {volume} {45}},\ \bibinfo {pages} {074006} (\bibinfo {year}
  {2012})}\BibitemShut {NoStop}%
\bibitem [{\citenamefont {Rybka}\ \emph {et~al.}(2016)\citenamefont {Rybka},
  \citenamefont {Ludwig}, \citenamefont {Schmalz}, \citenamefont {Knittel},
  \citenamefont {Brida},\ and\ \citenamefont {Leitenstorfer}}]{rybka2016sub}%
  \BibitemOpen
  \bibfield  {author} {\bibinfo {author} {\bibfnamefont {T.}~\bibnamefont
  {Rybka}}, \bibinfo {author} {\bibfnamefont {M.}~\bibnamefont {Ludwig}},
  \bibinfo {author} {\bibfnamefont {M.~F.}\ \bibnamefont {Schmalz}}, \bibinfo
  {author} {\bibfnamefont {V.}~\bibnamefont {Knittel}}, \bibinfo {author}
  {\bibfnamefont {D.}~\bibnamefont {Brida}},\ and\ \bibinfo {author}
  {\bibfnamefont {A.}~\bibnamefont {Leitenstorfer}},\ }\bibfield  {title}
  {\bibinfo {title} {Sub-cycle optical phase control of nanotunnelling in the
  single-electron regime},\ }\href@noop {} {\bibfield  {journal} {\bibinfo
  {journal} {Nature Photonics}\ }\textbf {\bibinfo {volume} {10}},\ \bibinfo
  {pages} {667} (\bibinfo {year} {2016})}\BibitemShut {NoStop}%
\bibitem [{\citenamefont {Ludwig}\ \emph {et~al.}(2020)\citenamefont {Ludwig},
  \citenamefont {Aguirregabiria}, \citenamefont {Ritzkowsky}, \citenamefont
  {Rybka}, \citenamefont {Marinica}, \citenamefont {Aizpurua}, \citenamefont
  {Borisov}, \citenamefont {Leitenstorfer},\ and\ \citenamefont
  {Brida}}]{ludwig2020sub}%
  \BibitemOpen
  \bibfield  {author} {\bibinfo {author} {\bibfnamefont {M.}~\bibnamefont
  {Ludwig}}, \bibinfo {author} {\bibfnamefont {G.}~\bibnamefont
  {Aguirregabiria}}, \bibinfo {author} {\bibfnamefont {F.}~\bibnamefont
  {Ritzkowsky}}, \bibinfo {author} {\bibfnamefont {T.}~\bibnamefont {Rybka}},
  \bibinfo {author} {\bibfnamefont {D.~C.}\ \bibnamefont {Marinica}}, \bibinfo
  {author} {\bibfnamefont {J.}~\bibnamefont {Aizpurua}}, \bibinfo {author}
  {\bibfnamefont {A.~G.}\ \bibnamefont {Borisov}}, \bibinfo {author}
  {\bibfnamefont {A.}~\bibnamefont {Leitenstorfer}},\ and\ \bibinfo {author}
  {\bibfnamefont {D.}~\bibnamefont {Brida}},\ }\bibfield  {title} {\bibinfo
  {title} {Sub-femtosecond electron transport in a nanoscale gap},\ }\href@noop
  {} {\bibfield  {journal} {\bibinfo  {journal} {Nature Physics}\ }\textbf
  {\bibinfo {volume} {16}},\ \bibinfo {pages} {341} (\bibinfo {year}
  {2020})}\BibitemShut {NoStop}%
\bibitem [{\citenamefont {Kim}\ \emph {et~al.}(2023)\citenamefont {Kim},
  \citenamefont {Garg}, \citenamefont {Mandal}, \citenamefont {Seiffert},
  \citenamefont {Fennel},\ and\ \citenamefont
  {Goulielmakis}}]{kim2023attosecond}%
  \BibitemOpen
  \bibfield  {author} {\bibinfo {author} {\bibfnamefont {H.}~\bibnamefont
  {Kim}}, \bibinfo {author} {\bibfnamefont {M.}~\bibnamefont {Garg}}, \bibinfo
  {author} {\bibfnamefont {S.}~\bibnamefont {Mandal}}, \bibinfo {author}
  {\bibfnamefont {L.}~\bibnamefont {Seiffert}}, \bibinfo {author}
  {\bibfnamefont {T.}~\bibnamefont {Fennel}},\ and\ \bibinfo {author}
  {\bibfnamefont {E.}~\bibnamefont {Goulielmakis}},\ }\bibfield  {title}
  {\bibinfo {title} {Attosecond field emission},\ }\href@noop {} {\bibfield
  {journal} {\bibinfo  {journal} {Nature}\ }\textbf {\bibinfo {volume} {613}},\
  \bibinfo {pages} {662} (\bibinfo {year} {2023})}\BibitemShut {NoStop}%
\bibitem [{\citenamefont {Arashida}\ \emph {et~al.}(2022)\citenamefont
  {Arashida}, \citenamefont {Mogi}, \citenamefont {Ishikawa}, \citenamefont
  {Igarashi}, \citenamefont {Hatanaka}, \citenamefont {Umeda}, \citenamefont
  {Peng}, \citenamefont {Yoshida}, \citenamefont {Takeuchi},\ and\
  \citenamefont {Shigekawa}}]{arashida2022subcycle}%
  \BibitemOpen
  \bibfield  {author} {\bibinfo {author} {\bibfnamefont {Y.}~\bibnamefont
  {Arashida}}, \bibinfo {author} {\bibfnamefont {H.}~\bibnamefont {Mogi}},
  \bibinfo {author} {\bibfnamefont {M.}~\bibnamefont {Ishikawa}}, \bibinfo
  {author} {\bibfnamefont {I.}~\bibnamefont {Igarashi}}, \bibinfo {author}
  {\bibfnamefont {A.}~\bibnamefont {Hatanaka}}, \bibinfo {author}
  {\bibfnamefont {N.}~\bibnamefont {Umeda}}, \bibinfo {author} {\bibfnamefont
  {J.}~\bibnamefont {Peng}}, \bibinfo {author} {\bibfnamefont {S.}~\bibnamefont
  {Yoshida}}, \bibinfo {author} {\bibfnamefont {O.}~\bibnamefont {Takeuchi}},\
  and\ \bibinfo {author} {\bibfnamefont {H.}~\bibnamefont {Shigekawa}},\
  }\bibfield  {title} {\bibinfo {title} {Subcycle mid-infrared
  electric-field-driven scanning tunneling microscopy with a time resolution
  higher than 30 fs},\ }\href@noop {} {\bibfield  {journal} {\bibinfo
  {journal} {ACS Photonics}\ }\textbf {\bibinfo {volume} {9}},\ \bibinfo
  {pages} {3156} (\bibinfo {year} {2022})}\BibitemShut {NoStop}%
\bibitem [{\citenamefont {Goulielmakis}\ \emph {et~al.}(2007)\citenamefont
  {Goulielmakis}, \citenamefont {Yakovlev}, \citenamefont {Cavalieri},
  \citenamefont {Uiberacker}, \citenamefont {Pervak}, \citenamefont
  {Apolonski}, \citenamefont {Kienberger}, \citenamefont {Kleineberg},\ and\
  \citenamefont {Krausz}}]{goulielmakis2007attosecond}%
  \BibitemOpen
  \bibfield  {author} {\bibinfo {author} {\bibfnamefont {E.}~\bibnamefont
  {Goulielmakis}}, \bibinfo {author} {\bibfnamefont {V.~S.}\ \bibnamefont
  {Yakovlev}}, \bibinfo {author} {\bibfnamefont {A.~L.}\ \bibnamefont
  {Cavalieri}}, \bibinfo {author} {\bibfnamefont {M.}~\bibnamefont
  {Uiberacker}}, \bibinfo {author} {\bibfnamefont {V.}~\bibnamefont {Pervak}},
  \bibinfo {author} {\bibfnamefont {A.}~\bibnamefont {Apolonski}}, \bibinfo
  {author} {\bibfnamefont {R.}~\bibnamefont {Kienberger}}, \bibinfo {author}
  {\bibfnamefont {U.}~\bibnamefont {Kleineberg}},\ and\ \bibinfo {author}
  {\bibfnamefont {F.}~\bibnamefont {Krausz}},\ }\bibfield  {title} {\bibinfo
  {title} {Attosecond control and measurement: lightwave electronics},\
  }\href@noop {} {\bibfield  {journal} {\bibinfo  {journal} {Science}\ }\textbf
  {\bibinfo {volume} {317}},\ \bibinfo {pages} {769} (\bibinfo {year}
  {2007})}\BibitemShut {NoStop}%
\bibitem [{\citenamefont {Bionta}\ \emph {et~al.}(2021)\citenamefont {Bionta},
  \citenamefont {Ritzkowsky}, \citenamefont {Turchetti}, \citenamefont {Yang},
  \citenamefont {Cattozzo~Mor}, \citenamefont {Putnam}, \citenamefont
  {K{\"a}rtner}, \citenamefont {Berggren},\ and\ \citenamefont
  {Keathley}}]{bionta2021chip}%
  \BibitemOpen
  \bibfield  {author} {\bibinfo {author} {\bibfnamefont {M.~R.}\ \bibnamefont
  {Bionta}}, \bibinfo {author} {\bibfnamefont {F.}~\bibnamefont {Ritzkowsky}},
  \bibinfo {author} {\bibfnamefont {M.}~\bibnamefont {Turchetti}}, \bibinfo
  {author} {\bibfnamefont {Y.}~\bibnamefont {Yang}}, \bibinfo {author}
  {\bibfnamefont {D.}~\bibnamefont {Cattozzo~Mor}}, \bibinfo {author}
  {\bibfnamefont {W.~P.}\ \bibnamefont {Putnam}}, \bibinfo {author}
  {\bibfnamefont {F.~X.}\ \bibnamefont {K{\"a}rtner}}, \bibinfo {author}
  {\bibfnamefont {K.~K.}\ \bibnamefont {Berggren}},\ and\ \bibinfo {author}
  {\bibfnamefont {P.~D.}\ \bibnamefont {Keathley}},\ }\bibfield  {title}
  {\bibinfo {title} {On-chip sampling of optical fields with attosecond
  resolution},\ }\href@noop {} {\bibfield  {journal} {\bibinfo  {journal}
  {Nature Photonics}\ }\textbf {\bibinfo {volume} {15}},\ \bibinfo {pages}
  {456} (\bibinfo {year} {2021})}\BibitemShut {NoStop}%
\bibitem [{\citenamefont {Herink}\ \emph {et~al.}(2012)\citenamefont {Herink},
  \citenamefont {Solli}, \citenamefont {Gulde},\ and\ \citenamefont
  {Ropers}}]{herink2012field}%
  \BibitemOpen
  \bibfield  {author} {\bibinfo {author} {\bibfnamefont {G.}~\bibnamefont
  {Herink}}, \bibinfo {author} {\bibfnamefont {D.~R.}\ \bibnamefont {Solli}},
  \bibinfo {author} {\bibfnamefont {M.}~\bibnamefont {Gulde}},\ and\ \bibinfo
  {author} {\bibfnamefont {C.}~\bibnamefont {Ropers}},\ }\bibfield  {title}
  {\bibinfo {title} {Field-driven photoemission from nanostructures quenches
  the quiver motion},\ }\href@noop {} {\bibfield  {journal} {\bibinfo
  {journal} {Nature}\ }\textbf {\bibinfo {volume} {483}},\ \bibinfo {pages}
  {190} (\bibinfo {year} {2012})}\BibitemShut {NoStop}%
\bibitem [{\citenamefont {Park}\ \emph {et~al.}(2012)\citenamefont {Park},
  \citenamefont {Piglosiewicz}, \citenamefont {Schmidt}, \citenamefont
  {Kollmann}, \citenamefont {Mascheck},\ and\ \citenamefont
  {Lienau}}]{park2012strong}%
  \BibitemOpen
  \bibfield  {author} {\bibinfo {author} {\bibfnamefont {D.~J.}\ \bibnamefont
  {Park}}, \bibinfo {author} {\bibfnamefont {B.}~\bibnamefont {Piglosiewicz}},
  \bibinfo {author} {\bibfnamefont {S.}~\bibnamefont {Schmidt}}, \bibinfo
  {author} {\bibfnamefont {H.}~\bibnamefont {Kollmann}}, \bibinfo {author}
  {\bibfnamefont {M.}~\bibnamefont {Mascheck}},\ and\ \bibinfo {author}
  {\bibfnamefont {C.}~\bibnamefont {Lienau}},\ }\bibfield  {title} {\bibinfo
  {title} {Strong field acceleration and steering of ultrafast electron pulses
  from a sharp metallic nanotip},\ }\href@noop {} {\bibfield  {journal}
  {\bibinfo  {journal} {Physical review letters}\ }\textbf {\bibinfo {volume}
  {109}},\ \bibinfo {pages} {244803} (\bibinfo {year} {2012})}\BibitemShut
  {NoStop}%
\bibitem [{\citenamefont {Putnam}\ \emph {et~al.}(2017)\citenamefont {Putnam},
  \citenamefont {Hobbs}, \citenamefont {Keathley}, \citenamefont {Berggren},\
  and\ \citenamefont {K{\"a}rtner}}]{putnam2017optical}%
  \BibitemOpen
  \bibfield  {author} {\bibinfo {author} {\bibfnamefont {W.~P.}\ \bibnamefont
  {Putnam}}, \bibinfo {author} {\bibfnamefont {R.~G.}\ \bibnamefont {Hobbs}},
  \bibinfo {author} {\bibfnamefont {P.~D.}\ \bibnamefont {Keathley}}, \bibinfo
  {author} {\bibfnamefont {K.~K.}\ \bibnamefont {Berggren}},\ and\ \bibinfo
  {author} {\bibfnamefont {F.~X.}\ \bibnamefont {K{\"a}rtner}},\ }\bibfield
  {title} {\bibinfo {title} {Optical-field-controlled photoemission from
  plasmonic nanoparticles},\ }\href@noop {} {\bibfield  {journal} {\bibinfo
  {journal} {nature physics}\ }\textbf {\bibinfo {volume} {13}},\ \bibinfo
  {pages} {335} (\bibinfo {year} {2017})}\BibitemShut {NoStop}%
\bibitem [{\citenamefont {Xiong}\ \emph {et~al.}(2020)\citenamefont {Xiong},
  \citenamefont {Zhou}, \citenamefont {Luo}, \citenamefont {Li}, \citenamefont
  {Bosman}, \citenamefont {Ang}, \citenamefont {Zhang},\ and\ \citenamefont
  {Wu}}]{xiong2020plasmon}%
  \BibitemOpen
  \bibfield  {author} {\bibinfo {author} {\bibfnamefont {X.}~\bibnamefont
  {Xiong}}, \bibinfo {author} {\bibfnamefont {Y.}~\bibnamefont {Zhou}},
  \bibinfo {author} {\bibfnamefont {Y.}~\bibnamefont {Luo}}, \bibinfo {author}
  {\bibfnamefont {X.}~\bibnamefont {Li}}, \bibinfo {author} {\bibfnamefont
  {M.}~\bibnamefont {Bosman}}, \bibinfo {author} {\bibfnamefont {L.~K.}\
  \bibnamefont {Ang}}, \bibinfo {author} {\bibfnamefont {P.}~\bibnamefont
  {Zhang}},\ and\ \bibinfo {author} {\bibfnamefont {L.}~\bibnamefont {Wu}},\
  }\bibfield  {title} {\bibinfo {title} {Plasmon-enhanced resonant
  photoemission using atomically thick dielectric coatings},\ }\href@noop {}
  {\bibfield  {journal} {\bibinfo  {journal} {ACS nano}\ }\textbf {\bibinfo
  {volume} {14}},\ \bibinfo {pages} {8806} (\bibinfo {year}
  {2020})}\BibitemShut {NoStop}%
\bibitem [{\citenamefont {Higuchi}\ \emph {et~al.}(2017)\citenamefont
  {Higuchi}, \citenamefont {Heide}, \citenamefont {Ullmann}, \citenamefont
  {Weber},\ and\ \citenamefont {Hommelhoff}}]{higuchi2017light}%
  \BibitemOpen
  \bibfield  {author} {\bibinfo {author} {\bibfnamefont {T.}~\bibnamefont
  {Higuchi}}, \bibinfo {author} {\bibfnamefont {C.}~\bibnamefont {Heide}},
  \bibinfo {author} {\bibfnamefont {K.}~\bibnamefont {Ullmann}}, \bibinfo
  {author} {\bibfnamefont {H.~B.}\ \bibnamefont {Weber}},\ and\ \bibinfo
  {author} {\bibfnamefont {P.}~\bibnamefont {Hommelhoff}},\ }\bibfield  {title}
  {\bibinfo {title} {Light-field-driven currents in graphene},\ }\href@noop {}
  {\bibfield  {journal} {\bibinfo  {journal} {Nature}\ }\textbf {\bibinfo
  {volume} {550}},\ \bibinfo {pages} {224} (\bibinfo {year}
  {2017})}\BibitemShut {NoStop}%
\bibitem [{\citenamefont {Son}\ \emph {et~al.}(2018)\citenamefont {Son},
  \citenamefont {Kim}, \citenamefont {Park}, \citenamefont {Lee}, \citenamefont
  {Park},\ and\ \citenamefont {Ahn}}]{son2018ultrafast}%
  \BibitemOpen
  \bibfield  {author} {\bibinfo {author} {\bibfnamefont {B.~H.}\ \bibnamefont
  {Son}}, \bibinfo {author} {\bibfnamefont {H.~S.}\ \bibnamefont {Kim}},
  \bibinfo {author} {\bibfnamefont {J.-Y.}\ \bibnamefont {Park}}, \bibinfo
  {author} {\bibfnamefont {S.}~\bibnamefont {Lee}}, \bibinfo {author}
  {\bibfnamefont {D.~J.}\ \bibnamefont {Park}},\ and\ \bibinfo {author}
  {\bibfnamefont {Y.~H.}\ \bibnamefont {Ahn}},\ }\bibfield  {title} {\bibinfo
  {title} {Ultrafast strong-field tunneling emission in graphene nanogaps},\
  }\href@noop {} {\bibfield  {journal} {\bibinfo  {journal} {ACS Photonics}\
  }\textbf {\bibinfo {volume} {5}},\ \bibinfo {pages} {3943} (\bibinfo {year}
  {2018})}\BibitemShut {NoStop}%
\bibitem [{\citenamefont {Zhou}\ \emph {et~al.}(2019)\citenamefont {Zhou},
  \citenamefont {Chen}, \citenamefont {Cole}, \citenamefont {Li}, \citenamefont
  {Chen}, \citenamefont {Li},\ and\ \citenamefont {Dai}}]{zhou2019ultrafast}%
  \BibitemOpen
  \bibfield  {author} {\bibinfo {author} {\bibfnamefont {S.}~\bibnamefont
  {Zhou}}, \bibinfo {author} {\bibfnamefont {K.}~\bibnamefont {Chen}}, \bibinfo
  {author} {\bibfnamefont {M.~T.}\ \bibnamefont {Cole}}, \bibinfo {author}
  {\bibfnamefont {Z.}~\bibnamefont {Li}}, \bibinfo {author} {\bibfnamefont
  {J.}~\bibnamefont {Chen}}, \bibinfo {author} {\bibfnamefont {C.}~\bibnamefont
  {Li}},\ and\ \bibinfo {author} {\bibfnamefont {Q.}~\bibnamefont {Dai}},\
  }\bibfield  {title} {\bibinfo {title} {Ultrafast field-emission electron
  sources based on nanomaterials},\ }\href@noop {} {\bibfield  {journal}
  {\bibinfo  {journal} {Advanced Materials}\ }\textbf {\bibinfo {volume}
  {31}},\ \bibinfo {pages} {1805845} (\bibinfo {year} {2019})}\BibitemShut
  {NoStop}%
\bibitem [{\citenamefont {Heide}\ \emph {et~al.}(2019)\citenamefont {Heide},
  \citenamefont {Boolakee}, \citenamefont {Higuchi}, \citenamefont {Weber},\
  and\ \citenamefont {Hommelhoff}}]{heide2019interaction}%
  \BibitemOpen
  \bibfield  {author} {\bibinfo {author} {\bibfnamefont {C.}~\bibnamefont
  {Heide}}, \bibinfo {author} {\bibfnamefont {T.}~\bibnamefont {Boolakee}},
  \bibinfo {author} {\bibfnamefont {T.}~\bibnamefont {Higuchi}}, \bibinfo
  {author} {\bibfnamefont {H.~B.}\ \bibnamefont {Weber}},\ and\ \bibinfo
  {author} {\bibfnamefont {P.}~\bibnamefont {Hommelhoff}},\ }\bibfield  {title}
  {\bibinfo {title} {Interaction of carrier envelope phase-stable laser pulses
  with graphene: the transition from the weak-field to the strong-field
  regime},\ }\href@noop {} {\bibfield  {journal} {\bibinfo  {journal} {New
  Journal of Physics}\ }\textbf {\bibinfo {volume} {21}},\ \bibinfo {pages}
  {045003} (\bibinfo {year} {2019})}\BibitemShut {NoStop}%
\bibitem [{\citenamefont {Sushko}\ \emph {et~al.}(2021)\citenamefont {Sushko},
  \citenamefont {De~Greve}, \citenamefont {Phillips}, \citenamefont {Urbaszek},
  \citenamefont {Joe}, \citenamefont {Watanabe}, \citenamefont {Taniguchi},
  \citenamefont {Efros}, \citenamefont {Hellberg}, \citenamefont {Park} \emph
  {et~al.}}]{sushko2021asymmetric}%
  \BibitemOpen
  \bibfield  {author} {\bibinfo {author} {\bibfnamefont {A.}~\bibnamefont
  {Sushko}}, \bibinfo {author} {\bibfnamefont {K.}~\bibnamefont {De~Greve}},
  \bibinfo {author} {\bibfnamefont {M.}~\bibnamefont {Phillips}}, \bibinfo
  {author} {\bibfnamefont {B.}~\bibnamefont {Urbaszek}}, \bibinfo {author}
  {\bibfnamefont {A.~Y.}\ \bibnamefont {Joe}}, \bibinfo {author} {\bibfnamefont
  {K.}~\bibnamefont {Watanabe}}, \bibinfo {author} {\bibfnamefont
  {T.}~\bibnamefont {Taniguchi}}, \bibinfo {author} {\bibfnamefont {A.~L.}\
  \bibnamefont {Efros}}, \bibinfo {author} {\bibfnamefont {C.~S.}\ \bibnamefont
  {Hellberg}}, \bibinfo {author} {\bibfnamefont {H.}~\bibnamefont {Park}},
  \emph {et~al.},\ }\bibfield  {title} {\bibinfo {title} {Asymmetric
  photoelectric effect: Auger-assisted hot hole photocurrents in transition
  metal dichalcogenides},\ }\href@noop {} {\bibfield  {journal} {\bibinfo
  {journal} {Nanophotonics}\ }\textbf {\bibinfo {volume} {10}},\ \bibinfo
  {pages} {105} (\bibinfo {year} {2021})}\BibitemShut {NoStop}%
\bibitem [{\citenamefont {Weiss}\ \emph {et~al.}(2012)\citenamefont {Weiss},
  \citenamefont {Zhou}, \citenamefont {Liao}, \citenamefont {Liu},
  \citenamefont {Jiang}, \citenamefont {Huang},\ and\ \citenamefont
  {Duan}}]{weiss2012graphene}%
  \BibitemOpen
  \bibfield  {author} {\bibinfo {author} {\bibfnamefont {N.~O.}\ \bibnamefont
  {Weiss}}, \bibinfo {author} {\bibfnamefont {H.}~\bibnamefont {Zhou}},
  \bibinfo {author} {\bibfnamefont {L.}~\bibnamefont {Liao}}, \bibinfo {author}
  {\bibfnamefont {Y.}~\bibnamefont {Liu}}, \bibinfo {author} {\bibfnamefont
  {S.}~\bibnamefont {Jiang}}, \bibinfo {author} {\bibfnamefont
  {Y.}~\bibnamefont {Huang}},\ and\ \bibinfo {author} {\bibfnamefont
  {X.}~\bibnamefont {Duan}},\ }\bibfield  {title} {\bibinfo {title} {Graphene:
  an emerging electronic material},\ }\href@noop {} {\bibfield  {journal}
  {\bibinfo  {journal} {Advanced materials}\ }\textbf {\bibinfo {volume}
  {24}},\ \bibinfo {pages} {5782} (\bibinfo {year} {2012})}\BibitemShut
  {NoStop}%
\bibitem [{\citenamefont {Novoselov}\ \emph {et~al.}(2004)\citenamefont
  {Novoselov}, \citenamefont {Geim}, \citenamefont {Morozov}, \citenamefont
  {Jiang}, \citenamefont {Zhang}, \citenamefont {Dubonos}, \citenamefont
  {Grigorieva},\ and\ \citenamefont {Firsov}}]{novoselov2004electric}%
  \BibitemOpen
  \bibfield  {author} {\bibinfo {author} {\bibfnamefont {K.~S.}\ \bibnamefont
  {Novoselov}}, \bibinfo {author} {\bibfnamefont {A.~K.}\ \bibnamefont {Geim}},
  \bibinfo {author} {\bibfnamefont {S.~V.}\ \bibnamefont {Morozov}}, \bibinfo
  {author} {\bibfnamefont {D.-e.}\ \bibnamefont {Jiang}}, \bibinfo {author}
  {\bibfnamefont {Y.}~\bibnamefont {Zhang}}, \bibinfo {author} {\bibfnamefont
  {S.~V.}\ \bibnamefont {Dubonos}}, \bibinfo {author} {\bibfnamefont {I.~V.}\
  \bibnamefont {Grigorieva}},\ and\ \bibinfo {author} {\bibfnamefont {A.~A.}\
  \bibnamefont {Firsov}},\ }\bibfield  {title} {\bibinfo {title} {Electric
  field effect in atomically thin carbon films},\ }\href@noop {} {\bibfield
  {journal} {\bibinfo  {journal} {science}\ }\textbf {\bibinfo {volume}
  {306}},\ \bibinfo {pages} {666} (\bibinfo {year} {2004})}\BibitemShut
  {NoStop}%
\bibitem [{\citenamefont {David}\ \emph {et~al.}(2014)\citenamefont {David},
  \citenamefont {Feldman}, \citenamefont {Mansfield}, \citenamefont {Lehman},\
  and\ \citenamefont {Singh}}]{david2014evaluating}%
  \BibitemOpen
  \bibfield  {author} {\bibinfo {author} {\bibfnamefont {L.}~\bibnamefont
  {David}}, \bibinfo {author} {\bibfnamefont {A.}~\bibnamefont {Feldman}},
  \bibinfo {author} {\bibfnamefont {E.}~\bibnamefont {Mansfield}}, \bibinfo
  {author} {\bibfnamefont {J.}~\bibnamefont {Lehman}},\ and\ \bibinfo {author}
  {\bibfnamefont {G.}~\bibnamefont {Singh}},\ }\bibfield  {title} {\bibinfo
  {title} {Evaluating the thermal damage resistance of graphene/carbon nanotube
  hybrid composite coatings},\ }\href@noop {} {\bibfield  {journal} {\bibinfo
  {journal} {Scientific reports}\ }\textbf {\bibinfo {volume} {4}},\ \bibinfo
  {pages} {1} (\bibinfo {year} {2014})}\BibitemShut {NoStop}%
\bibitem [{\citenamefont {Qin}\ \emph {et~al.}(2011)\citenamefont {Qin},
  \citenamefont {Wang}, \citenamefont {Xu}, \citenamefont {Li},\ and\
  \citenamefont {Forbes}}]{qin2011analytical}%
  \BibitemOpen
  \bibfield  {author} {\bibinfo {author} {\bibfnamefont {X.-Z.}\ \bibnamefont
  {Qin}}, \bibinfo {author} {\bibfnamefont {W.-L.}\ \bibnamefont {Wang}},
  \bibinfo {author} {\bibfnamefont {N.-S.}\ \bibnamefont {Xu}}, \bibinfo
  {author} {\bibfnamefont {Z.-B.}\ \bibnamefont {Li}},\ and\ \bibinfo {author}
  {\bibfnamefont {R.~G.}\ \bibnamefont {Forbes}},\ }\bibfield  {title}
  {\bibinfo {title} {Analytical treatment of cold field electron emission from
  a nanowall emitter, including quantum confinement effects},\ }\href@noop {}
  {\bibfield  {journal} {\bibinfo  {journal} {Proceedings of the Royal Society
  A: Mathematical, Physical and Engineering Sciences}\ }\textbf {\bibinfo
  {volume} {467}},\ \bibinfo {pages} {1029} (\bibinfo {year}
  {2011})}\BibitemShut {NoStop}%
\bibitem [{\citenamefont {Ang}\ \emph {et~al.}(2017)\citenamefont {Ang},
  \citenamefont {Liang},\ and\ \citenamefont {Ang}}]{ang2017theoretical}%
  \BibitemOpen
  \bibfield  {author} {\bibinfo {author} {\bibfnamefont {Y.}~\bibnamefont
  {Ang}}, \bibinfo {author} {\bibfnamefont {S.-J.}\ \bibnamefont {Liang}},\
  and\ \bibinfo {author} {\bibfnamefont {L.}~\bibnamefont {Ang}},\ }\bibfield
  {title} {\bibinfo {title} {Theoretical modeling of electron emission from
  graphene},\ }\href@noop {} {\bibfield  {journal} {\bibinfo  {journal} {MRS
  Bulletin}\ }\textbf {\bibinfo {volume} {42}},\ \bibinfo {pages} {505}
  (\bibinfo {year} {2017})}\BibitemShut {NoStop}%
\bibitem [{\citenamefont {Ang}\ \emph {et~al.}(2018)\citenamefont {Ang},
  \citenamefont {Yang},\ and\ \citenamefont {Ang}}]{ang2018universal}%
  \BibitemOpen
  \bibfield  {author} {\bibinfo {author} {\bibfnamefont {Y.~S.}\ \bibnamefont
  {Ang}}, \bibinfo {author} {\bibfnamefont {H.~Y.}\ \bibnamefont {Yang}},\ and\
  \bibinfo {author} {\bibfnamefont {L.}~\bibnamefont {Ang}},\ }\bibfield
  {title} {\bibinfo {title} {Universal scaling laws in schottky
  heterostructures based on two-dimensional materials},\ }\href@noop {}
  {\bibfield  {journal} {\bibinfo  {journal} {Physical review letters}\
  }\textbf {\bibinfo {volume} {121}},\ \bibinfo {pages} {056802} (\bibinfo
  {year} {2018})}\BibitemShut {NoStop}%
\bibitem [{\citenamefont {Ang}\ \emph {et~al.}(2019)\citenamefont {Ang},
  \citenamefont {Chen}, \citenamefont {Tan},\ and\ \citenamefont
  {Ang}}]{ang2019generalized}%
  \BibitemOpen
  \bibfield  {author} {\bibinfo {author} {\bibfnamefont {Y.~S.}\ \bibnamefont
  {Ang}}, \bibinfo {author} {\bibfnamefont {Y.}~\bibnamefont {Chen}}, \bibinfo
  {author} {\bibfnamefont {C.}~\bibnamefont {Tan}},\ and\ \bibinfo {author}
  {\bibfnamefont {L.}~\bibnamefont {Ang}},\ }\bibfield  {title} {\bibinfo
  {title} {Generalized high-energy thermionic electron injection at graphene
  interface},\ }\href@noop {} {\bibfield  {journal} {\bibinfo  {journal}
  {Physical Review Applied}\ }\textbf {\bibinfo {volume} {12}},\ \bibinfo
  {pages} {014057} (\bibinfo {year} {2019})}\BibitemShut {NoStop}%
\bibitem [{\citenamefont {Ang}\ \emph {et~al.}(2021)\citenamefont {Ang},
  \citenamefont {Cao},\ and\ \citenamefont {Ang}}]{ang2021physics}%
  \BibitemOpen
  \bibfield  {author} {\bibinfo {author} {\bibfnamefont {Y.~S.}\ \bibnamefont
  {Ang}}, \bibinfo {author} {\bibfnamefont {L.}~\bibnamefont {Cao}},\ and\
  \bibinfo {author} {\bibfnamefont {L.~K.}\ \bibnamefont {Ang}},\ }\bibfield
  {title} {\bibinfo {title} {Physics of electron emission and injection in
  two-dimensional materials: Theory and simulation},\ }\href@noop {} {\bibfield
   {journal} {\bibinfo  {journal} {InfoMat}\ }\textbf {\bibinfo {volume} {3}},\
  \bibinfo {pages} {502} (\bibinfo {year} {2021})}\BibitemShut {NoStop}%
\bibitem [{\citenamefont {Chan}\ \emph {et~al.}(2021)\citenamefont {Chan},
  \citenamefont {Ang},\ and\ \citenamefont {Ang}}]{chan2021thermal}%
  \BibitemOpen
  \bibfield  {author} {\bibinfo {author} {\bibfnamefont {W.~J.}\ \bibnamefont
  {Chan}}, \bibinfo {author} {\bibfnamefont {Y.~S.}\ \bibnamefont {Ang}},\ and\
  \bibinfo {author} {\bibfnamefont {L.}~\bibnamefont {Ang}},\ }\bibfield
  {title} {\bibinfo {title} {Thermal-field electron emission from
  three-dimensional dirac and weyl semimetals},\ }\href@noop {} {\bibfield
  {journal} {\bibinfo  {journal} {Physical Review B}\ }\textbf {\bibinfo
  {volume} {104}},\ \bibinfo {pages} {245420} (\bibinfo {year}
  {2021})}\BibitemShut {NoStop}%
\bibitem [{\citenamefont {Chan}\ \emph {et~al.}(2022)\citenamefont {Chan},
  \citenamefont {Chua}, \citenamefont {Ang},\ and\ \citenamefont
  {Ang}}]{chan2022field}%
  \BibitemOpen
  \bibfield  {author} {\bibinfo {author} {\bibfnamefont {W.~J.}\ \bibnamefont
  {Chan}}, \bibinfo {author} {\bibfnamefont {C.}~\bibnamefont {Chua}}, \bibinfo
  {author} {\bibfnamefont {Y.~S.}\ \bibnamefont {Ang}},\ and\ \bibinfo {author}
  {\bibfnamefont {L.~K.}\ \bibnamefont {Ang}},\ }\bibfield  {title} {\bibinfo
  {title} {Field emission in emerging two-dimensional and topological
  materials: A perspective},\ }\href@noop {} {\bibfield  {journal} {\bibinfo
  {journal} {IEEE Transactions on Plasma Science}\ } (\bibinfo {year}
  {2022})}\BibitemShut {NoStop}%
\bibitem [{\citenamefont {Ang}\ \emph {et~al.}(2023)\citenamefont {Ang},
  \citenamefont {Ang},\ and\ \citenamefont {Lee}}]{ang2023universal}%
  \BibitemOpen
  \bibfield  {author} {\bibinfo {author} {\bibfnamefont {L.~K.}\ \bibnamefont
  {Ang}}, \bibinfo {author} {\bibfnamefont {Y.~S.}\ \bibnamefont {Ang}},\ and\
  \bibinfo {author} {\bibfnamefont {C.~H.}\ \bibnamefont {Lee}},\ }\bibfield
  {title} {\bibinfo {title} {Universal model for electron thermal-field
  emission from two-dimensional semimetals},\ }\href@noop {} {\bibfield
  {journal} {\bibinfo  {journal} {Physics of Plasmas}\ }\textbf {\bibinfo
  {volume} {30}},\ \bibinfo {pages} {033103} (\bibinfo {year}
  {2023})}\BibitemShut {NoStop}%
\bibitem [{\citenamefont {Fowler}\ and\ \citenamefont
  {Nordheim}(1928)}]{fowler1928electron}%
  \BibitemOpen
  \bibfield  {author} {\bibinfo {author} {\bibfnamefont {R.~H.}\ \bibnamefont
  {Fowler}}\ and\ \bibinfo {author} {\bibfnamefont {L.}~\bibnamefont
  {Nordheim}},\ }\bibfield  {title} {\bibinfo {title} {Electron emission in
  intense electric fields},\ }\href@noop {} {\bibfield  {journal} {\bibinfo
  {journal} {Proceedings of the Royal Society of London. Series A, Containing
  Papers of a Mathematical and Physical Character}\ }\textbf {\bibinfo {volume}
  {119}},\ \bibinfo {pages} {173} (\bibinfo {year} {1928})}\BibitemShut
  {NoStop}%
\bibitem [{\citenamefont {Chua}\ \emph {et~al.}(2021)\citenamefont {Chua},
  \citenamefont {Kee}, \citenamefont {Ang},\ and\ \citenamefont
  {Ang}}]{chua2021absence}%
  \BibitemOpen
  \bibfield  {author} {\bibinfo {author} {\bibfnamefont {C.}~\bibnamefont
  {Chua}}, \bibinfo {author} {\bibfnamefont {C.~Y.}\ \bibnamefont {Kee}},
  \bibinfo {author} {\bibfnamefont {Y.~S.}\ \bibnamefont {Ang}},\ and\ \bibinfo
  {author} {\bibfnamefont {L.}~\bibnamefont {Ang}},\ }\bibfield  {title}
  {\bibinfo {title} {Absence of space-charge-limited current in unconventional
  field emission},\ }\href@noop {} {\bibfield  {journal} {\bibinfo  {journal}
  {Physical Review Applied}\ }\textbf {\bibinfo {volume} {16}},\ \bibinfo
  {pages} {064025} (\bibinfo {year} {2021})}\BibitemShut {NoStop}%
\bibitem [{\citenamefont {Meshkov}(1986)}]{meshkov1986tunneling}%
  \BibitemOpen
  \bibfield  {author} {\bibinfo {author} {\bibfnamefont {S.}~\bibnamefont
  {Meshkov}},\ }\bibfield  {title} {\bibinfo {title} {Tunneling of electrons
  from a two-dimensional channel into the bulk},\ }\href@noop {} {\bibfield
  {journal} {\bibinfo  {journal} {Zh. Eksp. Teor. Fiz}\ }\textbf {\bibinfo
  {volume} {91}},\ \bibinfo {pages} {198} (\bibinfo {year} {1986})}\BibitemShut
  {NoStop}%
\bibitem [{\citenamefont {Chandni}\ \emph {et~al.}(2016)\citenamefont
  {Chandni}, \citenamefont {Watanabe}, \citenamefont {Taniguchi},\ and\
  \citenamefont {Eisenstein}}]{chandni2016signatures}%
  \BibitemOpen
  \bibfield  {author} {\bibinfo {author} {\bibfnamefont {U.}~\bibnamefont
  {Chandni}}, \bibinfo {author} {\bibfnamefont {K.}~\bibnamefont {Watanabe}},
  \bibinfo {author} {\bibfnamefont {T.}~\bibnamefont {Taniguchi}},\ and\
  \bibinfo {author} {\bibfnamefont {J.~P.}\ \bibnamefont {Eisenstein}},\
  }\bibfield  {title} {\bibinfo {title} {Signatures of phonon and
  defect-assisted tunneling in planar metal--hexagonal boron nitride--graphene
  junctions},\ }\href@noop {} {\bibfield  {journal} {\bibinfo  {journal} {Nano
  letters}\ }\textbf {\bibinfo {volume} {16}},\ \bibinfo {pages} {7982}
  (\bibinfo {year} {2016})}\BibitemShut {NoStop}%
\bibitem [{\citenamefont {Huang}\ \emph {et~al.}(2006)\citenamefont {Huang},
  \citenamefont {Lau}, \citenamefont {Yang},\ and\ \citenamefont
  {Yu}}]{huang2006local}%
  \BibitemOpen
  \bibfield  {author} {\bibinfo {author} {\bibfnamefont {L.}~\bibnamefont
  {Huang}}, \bibinfo {author} {\bibfnamefont {S.~P.}\ \bibnamefont {Lau}},
  \bibinfo {author} {\bibfnamefont {H.}~\bibnamefont {Yang}},\ and\ \bibinfo
  {author} {\bibfnamefont {S.}~\bibnamefont {Yu}},\ }\bibfield  {title}
  {\bibinfo {title} {Local measurement of secondary electron emission from
  zno-coated carbon nanotubes},\ }\href@noop {} {\bibfield  {journal} {\bibinfo
   {journal} {Nanotechnology}\ }\textbf {\bibinfo {volume} {17}},\ \bibinfo
  {pages} {1564} (\bibinfo {year} {2006})}\BibitemShut {NoStop}%
\bibitem [{\citenamefont {Echternkamp}\ \emph {et~al.}(2016)\citenamefont
  {Echternkamp}, \citenamefont {Herink}, \citenamefont {Yalunin}, \citenamefont
  {Rademann}, \citenamefont {Sch{\"a}fer},\ and\ \citenamefont
  {Ropers}}]{echternkamp2016strong}%
  \BibitemOpen
  \bibfield  {author} {\bibinfo {author} {\bibfnamefont {K.}~\bibnamefont
  {Echternkamp}}, \bibinfo {author} {\bibfnamefont {G.}~\bibnamefont {Herink}},
  \bibinfo {author} {\bibfnamefont {S.~V.}\ \bibnamefont {Yalunin}}, \bibinfo
  {author} {\bibfnamefont {K.}~\bibnamefont {Rademann}}, \bibinfo {author}
  {\bibfnamefont {S.}~\bibnamefont {Sch{\"a}fer}},\ and\ \bibinfo {author}
  {\bibfnamefont {C.}~\bibnamefont {Ropers}},\ }\bibfield  {title} {\bibinfo
  {title} {Strong-field photoemission in nanotip near-fields: from quiver to
  sub-cycle electron dynamics},\ }\href@noop {} {\bibfield  {journal} {\bibinfo
   {journal} {Applied Physics B}\ }\textbf {\bibinfo {volume} {122}},\ \bibinfo
  {pages} {1} (\bibinfo {year} {2016})}\BibitemShut {NoStop}%
\bibitem [{\citenamefont {Jensen}(2007)}]{jensen2007general}%
  \BibitemOpen
  \bibfield  {author} {\bibinfo {author} {\bibfnamefont {K.~L.}\ \bibnamefont
  {Jensen}},\ }\bibfield  {title} {\bibinfo {title} {General formulation of
  thermal, field, and photoinduced electron emission},\ }\href@noop {}
  {\bibfield  {journal} {\bibinfo  {journal} {Journal of Applied Physics}\
  }\textbf {\bibinfo {volume} {102}},\ \bibinfo {pages} {024911} (\bibinfo
  {year} {2007})}\BibitemShut {NoStop}%
\end{thebibliography}
\end{document}